\documentclass[10pt,
							 twocolumn,
							 superscriptaddress,
							 english,
							 prl,
							 showpacs,
							 floatfix,
							 aps
							]{revtex4-1}
\usepackage[utf8]{inputenc}
\usepackage{amsmath}
\usepackage{amssymb}
\usepackage{graphicx}
\usepackage{xspace}

\makeatletter
\@ifundefined{textcolor}{}
{%
 \definecolor{BLACK}{gray}{0}
 \definecolor{WHITE}{gray}{1}
 \definecolor{RED}{rgb}{1,0,0}
 \definecolor{GREEN}{rgb}{0,1,0}
 \definecolor{BLUE}{rgb}{0,0,1}
 \definecolor{CYAN}{cmyk}{1,0,0,0}
 \definecolor{MAGENTA}{cmyk}{0,1,0,0}
 \definecolor{YELLOW}{cmyk}{0,0,1,0}
}

\usepackage{soul}
\usepackage{braket}
\usepackage[backref=none,
bookmarksnumbered=true,
bookmarks=true,
bookmarksopen=true,
colorlinks=true,
citecolor=blue,
linkcolor=blue,
anchorcolor=green,
urlcolor=blue,unicode=false]{hyperref}

\renewcommand{\v}[1]{\ensuremath{\mathbf{#1}}} 
 
\let\baraccent=\= 
\renewcommand{\=}[1]{\stackrel{#1}{=}} 

\newcommand{\unitspace}{~}

\newcommand{\didv}{\ensuremath{\mathrm{d}I/\mathrm{d}V}\xspace}

\newcommand{\DeltaT}{\ensuremath{\Delta_\mathrm{t}}\xspace}

\newcommand{\Fig}[1]{Fig.\unitspace\ref{fig:#1}}
\newcommand{\Figure}[1]{Figure\unitspace\ref{fig:#1}}

\DeclareMathOperator{\upm}{\unitspace\mathrm{pm}}

\DeclareMathOperator{\umV}{\unitspace\mathrm{mV}}
\DeclareMathOperator{\umuV}{\unitspace\mathrm{\mu V}}

\DeclareMathOperator{\umueV}{\unitspace\mathrm{\mu eV}}

\DeclareMathOperator{\upA}{\unitspace\mathrm{pA}}

\DeclareMathOperator{\uK}{\unitspace\mathrm{K}}

\makeatother

\usepackage{babel}

\begin{document}

\title{Tuning the coupling of an individual magnetic impurity to a superconductor:\\ quantum phase transition and transport}

\author{La\"etitia Farinacci}
\affiliation{\mbox{Fachbereich Physik, Freie Universit\"at Berlin, Arnimallee 14, 14195 Berlin, Germany}}

\author{Gelavizh Ahmadi}
\affiliation{\mbox{Fachbereich Physik, Freie Universit\"at Berlin, Arnimallee 14, 14195 Berlin, Germany}}

\author{Ga\"el Reecht}
\affiliation{\mbox{Fachbereich Physik, Freie Universit\"at Berlin, Arnimallee 14, 14195 Berlin, Germany}}

\author{Michael Ruby}
\affiliation{\mbox{Fachbereich Physik, Freie Universit\"at Berlin, Arnimallee 14, 14195 Berlin, Germany}}

\author{Nils Bogdanoff}
\affiliation{\mbox{Fachbereich Physik, Freie Universit\"at Berlin, Arnimallee 14, 14195 Berlin, Germany}}

\author{Olof Peters}
\affiliation{\mbox{Fachbereich Physik, Freie Universit\"at Berlin, Arnimallee 14, 14195 Berlin, Germany}}

\author{Benjamin W. Heinrich}
\affiliation{\mbox{Fachbereich Physik, Freie Universit\"at Berlin, Arnimallee 14, 14195 Berlin, Germany}}

\author{Felix von Oppen}
\affiliation{\mbox{Dahlem Center for Complex Quantum Systems and Fachbereich Physik, Freie Universit\"at Berlin, 14195 Berlin, Germany}}

\author{Katharina J. Franke}
\affiliation{\mbox{Fachbereich Physik, Freie Universit\"at Berlin, Arnimallee 14, 14195 Berlin, Germany}}

\date{\today}

\begin{abstract}

The exchange scattering at magnetic adsorbates on superconductors gives rise to Yu-Shiba-Rusinov (YSR) bound states. Depending on the strength of the exchange coupling, the magnetic moment perturbs the Cooper pair condensate only weakly, resulting in a free-spin ground state, or binds a quasiparticle in its vicinity, leading to a (partially) screened spin state. Here, we use the flexibility of Fe-porphin (FeP) molecules adsorbed on a Pb(111) surface to reversibly and continuously tune between these distinct ground states. We find that the FeP moment is screened in the pristine adsorption state. Approaching the tip of a scanning tunneling microscope, we exert a sufficiently strong attractive force to tune the molecule through the quantum phase transition into the free-spin state. We ascertain and characterize the transition by investigating the transport processes as function of tip-molecule distance, exciting the YSR states by single-electron tunneling as well as (multiple) Andreev reflections. 

\end{abstract}

\pacs{%
			} 
\maketitle 

\begin{figure}[tb]
	\includegraphics[width=\columnwidth]{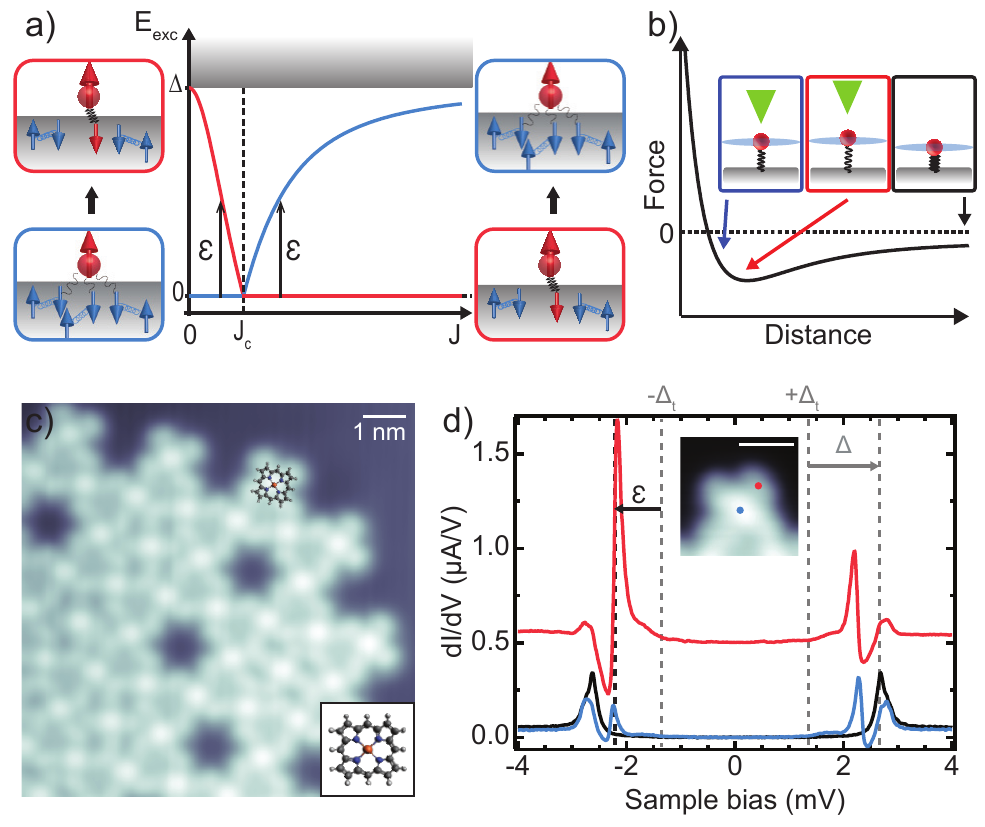}
	\caption{a) Schematic dependence of ground and excited states of a magnetic adatom on a superconducting substrate on exchange coupling strength $J$. At weak coupling, the ground state is a free-spin state, while the excited state has a quasiparticle bound to the adatom. At a critical coupling $J_c$, the roles of ground and excited states reverse. b) Sketch of the forces acting between tip and molecule vs.\ tip--molecule distance. At large distances, attractive forces pull the molecule away from the surface. At closer distance, repulsive forces push the molecule back towards the surface. c) STM topography image of a FeP island ($V=-45$~mV, $I=50$~pA) with a molecular model (inset). d) \didv spectra acquired with a Pb tip above bare Pb(111) (black), center (blue) and ligand (red, offset for clarity) of a molecule.}
	\label{fig:1}
\end{figure}

The exchange coupling of magnetic impurities to a superconductor induces localized Yu-Shiba-Rusinov (YSR) bound states \cite{Yu1965,Shiba1968,Rusinov1969}. Even for a single impurity, the local nature of the superconducting ground state depends qualitatively on the strength of the exchange coupling $J$ (\Fig{1}a). For weak coupling, the bound subgap state remains unoccupied, the superconducting ground state fully paired, and the impurity spin free. At strong coupling, the bound state becomes occupied and the superconducting ground state involves an unpaired electron which (partially) screens the impurity spin \cite{Sakurai1970, Matsuura1977, Sakai1993, Balatsky2006, Heinrich2018}. The level crossing between these states at a critical coupling $J_c$, commonly referred to as a quantum phase transition, is protected by fermion-parity conservation. As states of different fermion parity exchange roles at the transition, the level crossing is most immediately reflected in the single-particle addition spectrum. A discontinuity in the corresponding spectral weights as well as an abrupt change in the screened impurity spin make this a first-order transition. When including effects of self-consistency, additional discontinuities are predicted in the bound state energy and the local order parameter \cite{Salkola1997, Flatte1997, Meng2015}.

Experimental probes of magnetic-impurity physics use quantum dots or magnetic adatoms. For quantum dots, the exchange coupling $J$ can be controlled electrically.  Superconductor (SC)--quantum dot (QD)--SC junctions then provide indirect evidence for the quantum phase transition through a 0--$\pi$ transition of the Josephson current \cite{Maurand2012, Delagrange2015, Delagrange2016}. More immediate spectroscopic evidence emerges from measurements on asymmetric SC--QD--normal metal (N) junctions \cite{Deacon2010,Lee2014, Lee2017}. Typically obtained in the Coulomb-blockade regime at mK temperatures, the spectra are dominated by Andreev reflections at the junction and are thus insensitive to the spectral weights of the (single-particle) addition spectrum. 

In contrast, probing magnetic adatoms with a scanning tunneling microscope (STM) tip readily provides access to both, single-electron and Cooper-pair (Andreev) tunneling by controlling the tunnel gap and varying the junction conductance over several orders of magnitude~\cite{Ruby2015}. In the regime of single-electron tunneling, STM experiments measure the single-particle addition spectrum and thus provide crucial information about the quantum phase transition. In particular, tunneling spectroscopy (STS) provides access to the asymmetry between electron- and hole-like YSR excitations ~\cite{Ruby2015, Yazdani1997, Ji2008, Franke2011, Menard2015, Ruby2016, Choi2017, Cornils2017, Ruby2015chains} which changes abruptly at the transition.  

STM experiments are thus particularly well suited to probe the first-order nature of the transition. However, earlier experiments \cite{Franke2011, Hatter2015, Hatter2017} could only access a discrete set of exchange couplings $J$ determined by the adatom's adsorption site. Here, we use the flexibility of a single molecule adsorbed on superconducting Pb(111) to modify the molecule's interaction with the surface and tune the system continuously through the quantum phase transition. The superconducting tip exerts a mechanical force on the molecule which can be approximated by a Lennard-Jones potential (\Fig{1}b). At large tip--molecule distances, the force is attractive and the molecule is lifted from the surface. This modifies characteristic parameters in the junction and in particular reduces the exchange coupling~\cite{Heinrich2015, Hiraoka2017,Brand2018}. (For a discussion of the influence of other parameters, see supplementary material~\cite{SM}.) As the tip approaches, repulsive forces eventually dominate, push the molecule back towards the surface, and  increase the exchange coupling again. When combined with detailed conductance measurements to resolve the different tunneling processes, this technique allows for an in-depth analysis of the quantum phase transition.

We use Fe(II)-porphin (FeP) molecules deposited on a clean Pb(111) surface from a powder of Fe(III)-porphin-chloride (FeP-Cl). These molecules lose their Cl ligand by deposition below room temperature followed by annealing to 370$\uK$ (see supplementary material~\cite{SM}). STM experiments at a temperature of 1.6$\uK$ reveal the formation of well-ordered islands, as shown in \Fig{1}c. The individual molecules are identified by their clover shape with a bright protrusion at the Fe center. 

The superconducting Pb tips which we use to probe the YSR states of the molecule provide an effective energy resolution beyond the Fermi-Dirac limit. The measured signal is a convolution of the tip's Bardeen-Cooper-Schrieffer (BCS) density of states with the substrate density of states. Apart from peak intensity changes, the convolution shifts all spectral features of the substrate by the tip's superconducting energy gap $\DeltaT$. 
The BCS peaks of the substrate are thus observed at bias voltages of $\pm(\Delta+\DeltaT)/e=\pm2.65\umV$ \cite{RubyPb15}, where $\Delta$ denotes the gap of the substrate. Inside the gap, we find one pair of YSR states at $\pm2.2\umV$. These are resolved both on the Fe center and the ligand (\Fig{1}d) so that both positions can be employed to investigate the tip-induced forces and the quantum phase transition.

\begin{figure}[tb]
	\includegraphics[width=\columnwidth]{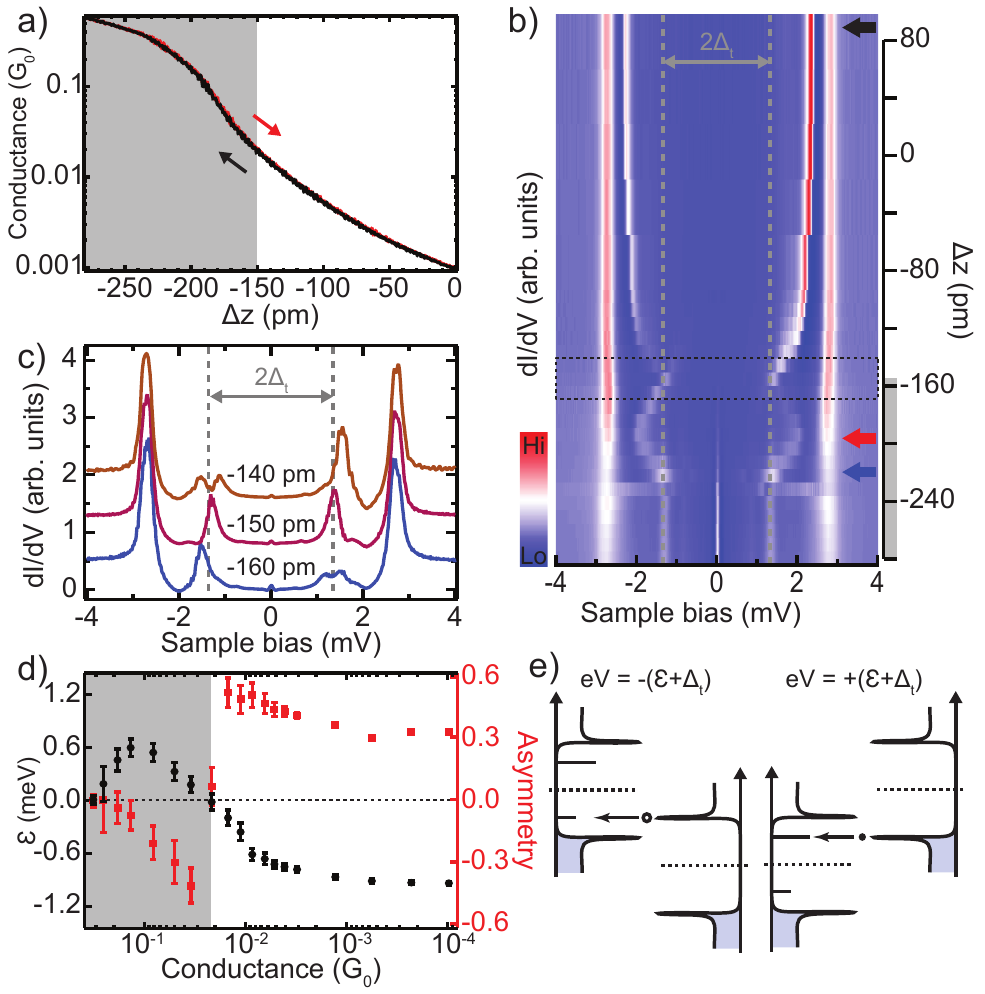}
	\caption{Tip approach above Fe center of a FeP molecule (for more data see~\cite{SM}). a) Junction conductance (in $G_0=2e^2/h$) vs.\ tip offset $\Delta z$. b) 2D false-color plot of \didv spectra recorded at various tip-sample distances,  normalized to their conductances at $5$~mV. The spectra were acquired after opening the feedback at $V=5$~mV, $I=200\upA$ and subsequently varying the tip height by an offset $\Delta z$ (lock-in modulation $V_\mathrm{rms}=15\umuV$ ). c) Spectra before ($\Delta z = -140$~pm, brown trace), at ($\Delta z=-150$~pm, claret), and after ($\Delta z =-160$~pm, blue) the quantum phase transition (offset for clarity). d) Extracted YSR energies and intensity asymmetries vs.\ junction conductance. Around $G=0.02 \times G_0$ ($\Delta z=-150$~pm), the YSR state energy crosses zero and the asymmetry changes sign, indicating the quantum phase transition. e) Single-electron tunneling processes for YSR resonances at positive and negative bias voltages.}
	\label{fig:2}
\end{figure}

We first characterize the junction conductance as the tip approaches the Fe center. We measure the tip approach $\Delta z$ from a set point at $V=5$~mV and $I=200\upA$. As shown in \Fig{2}a, the conductance first increases exponentially with $\Delta z$ as expected for a tunnel junction. (Small deviations are due to nonlinearities in the I-V converter at small current densities.)  A superexponential increase beyond $\Delta z = -150$~pm and subsequent leveling off indicate the transition to the contact regime between tip and molecule. Contact formation is fully reversible with identical approach and retraction curves, enabling precise tuning of the junction conductance. 
 
As described above, the strength of the exchange coupling and hence the binding energy of the YSR states depend sensitively on the tip approach $\Delta z$. As the tip approaches the Fe center, we observe that the YSR states shift deeper into the superconducting gap (\Fig{2}b). At $\Delta z = -150\upm$, the YSR resonance occurs at a bias voltage equal to the tip`s gap parameter, \textit{i.e.}, at zero energy (see zoom in \Fig{2}c). Approaching the tip further, the YSR states shift back towards the superconducting gap edge before eventually reversing at $\Delta z=-190\upm$ (red arrow in \Fig{2}b) and reaching zero energy for a second time (blue arrow in \Fig{2}b). Beyond this point, at even closer tip--molecule distances, the spectra are dominated by peaks at zero bias and $\DeltaT$. 

Individual \didv spectra do not allow for a complete identification of the binding energy of the YSR state since the same excitation energy may signify a screened or a free-spin ground state (see \Fig{1}a). However, the observed shift of the YSR states deeper into the superconducting energy gap upon approach allows for an unambiguous assignment. As attractive tip--molecule forces initially lift the Fe center from the surface and weaken the exchange coupling $J$, the pristine system must be in the strong-coupling regime with a screened-spin ground state and a negative YSR energy. 

We observe asymmetric intensities of the electron- and hole-like YSR resonances at positive and negative voltages, respectively. This asymmetry is a consequence of the potential scattering by the magnetic impurity. At large tip-molecule distances, the electron-like excitation has more spectral weight than the hole-like excitation. The relative strength changes with tip approach and accompanying YSR energy shift. In \Fig{2}d, we plot the binding energy $\epsilon$ of the YSR state and the asymmetry $(I_+-I_-)/(I_++I_-)$ as a function of junction conductance, with $I_{+/-}$ being the YSR intensities at positive and negative bias voltages, respectively. While the binding energy passes smoothly through zero, the zero-energy crossing is associated with an abrupt change in sign of the asymmetry. 

These observations provide detailed evidence for a first-order quantum phase transition from a screened-spin to a free-spin ground state as the tip--molecule distance is reduced. For sufficiently weak tunneling, the current is dominated by single-electron tunneling (sketches in \Fig{2}e)~\cite{Ruby2015}. Hence, we can relate the observed intensities of the electron- and hole-like resonances to the local weights of the electron and hole wave functions of the YSR bound state (see~\cite{SM} for additional details). Exciting the system out of the screened-spin YSR state annihilates a bound quasiparticle. For single-electron tunneling, this process involves $\gamma_0= u c_\alpha-v c^\dagger_\beta$, where $\gamma_0$ ($c_{\alpha}$) is the annihilation operator of the bound quasiparticle (electron with spin $\alpha$). Correspondingly, the intensity of the electron-like excitation at positive bias voltages is given by $|v|^2$, as the bound electron combines with the tunneling electron to form a Cooper pair. Similarly, the hole-like excitation at negative bias is given by $|u|^2$, reflecting a bound electron tunneling out into the tip. The roles of $u$ and $v$ are reversed when exciting the system out of the free-spin ground state and creating a bound quasiparticle, as described by $\gamma^\dagger_0 = u^*c^\dagger_\alpha-v^*c_\beta$. Now, electron tunneling at positive bias occupies the bound state and involves $|u|^2$, while tunneling at negative bias breaks a Cooper pair and involves $|v|^2$. The abrupt change in the asymmetry at a tip approach of $\Delta z=-150\upm$, where the YSR state crosses zero energy, is thus a hallmark of the first-order quantum phase transition. We also note that the gradual increase of the asymmetry before the quantum phase transition is in agreement with a decrease in the exchange coupling~\cite{Salkola1997, Bauer2013}. 

Self-consistency causes additional discontinuities in the bound-state energy and the local order parameter near the impurity \cite{Salkola1997,Flatte1997,Meng2015}. We do not find indications of these discontinuities. In the supplementary material~\cite{SM}, we derive analytical estimates for their magnitude. The discontinuity of the local gap parameter is substantial, of order $\delta\Delta\sim-\Delta/\ln(\omega_D/\Delta)$ within a few Fermi wavelengths of the impurity ($\omega_D$ is the Debye frequency), but cannot be directly probed by single-electron tunneling. The latter probes the bound state energy, whose jump is of order $\delta\epsilon\sim \Delta^2/[E_F\ln(\omega_D/\Delta)]\sim 10^{-2}\umueV$. This is several orders of magnitude below our experimental resolution of order $\sim 100 \umueV$. Instead of a discontinuity, we observe that the asymmetry vanishes right at the transition. This suggests that both $|u|^2$ and $|v|^2$ are probed simultaneously as is natural for a bound state whose energy vanishes within the resolution.

\begin{figure}[tb]
	\includegraphics[width=\columnwidth]{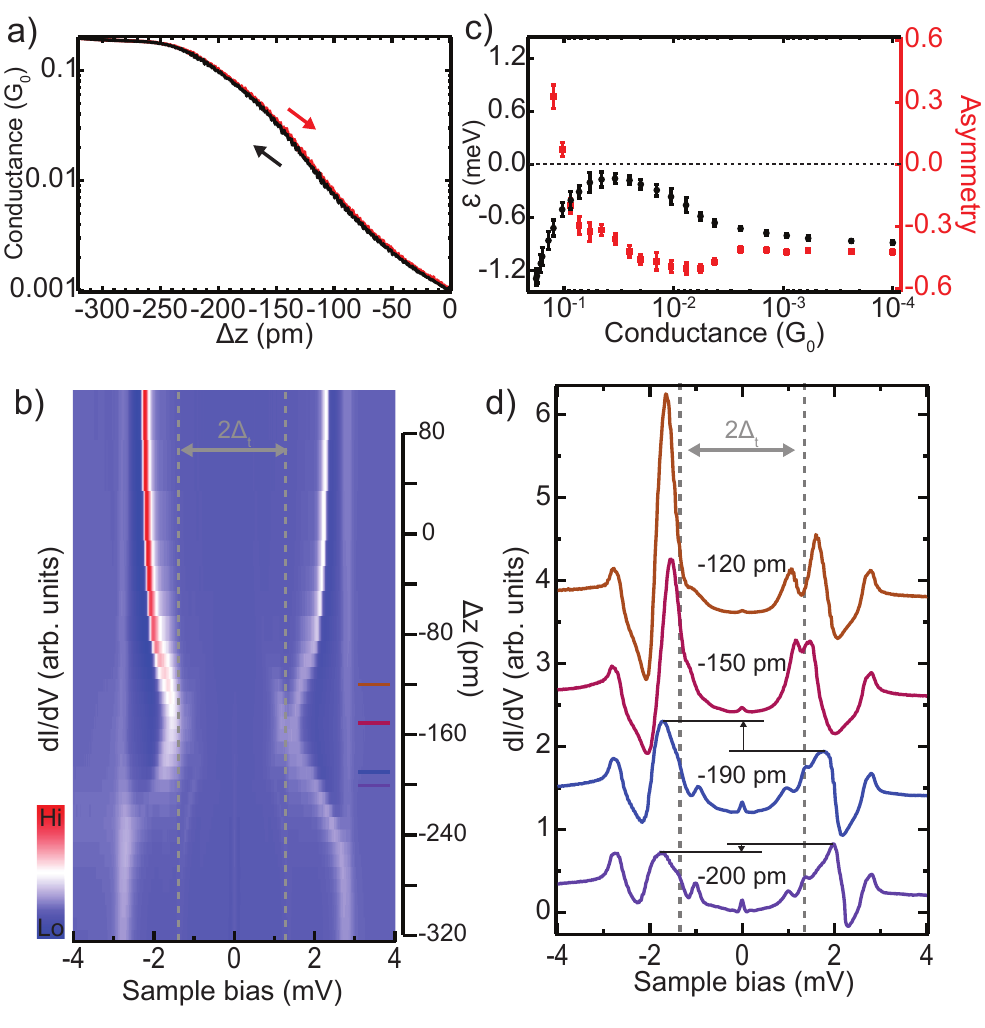}
	\caption{Tip approach above Fe ligand of FeP molecule (see \Fig{1}b) (for more data see~\cite{SM}). a) Evolution of conductance with tip offset $\Delta z$. b) 2D false-color plot of \didv spectra recorded at various tip-sample distances,  normalized to their conductances at $5$~mV. c) Extracted YSR energies and intensity asymmetries vs.\ junction conductance. d) Four spectra of this approach set (offset for clarity). The YSR state does not reach zero energy and the asymmetry remains positive during most of the approach, see the brown, claret, and blue spectra in panel d). The asymmetry reverses only around $G=0.1 \times G_0$ where Andreev reflections become dominant, 
see blue and purple spectra in panel d.}
	\label{fig:3}
\end{figure}

Next, we place the tip above the ligand. The $G-\Delta z$ curves exhibit the same stability and reversibility as on the center and allow for probing both the tunneling and the contact regime (\Fig{3}a). At large tip-molecule distances, we find YSR states at the same bias voltages as above the Fe center, but with reversed asymmetry (\Fig{1}d). The \didv spectra at different junction conductances (\Fig{3}b) also show the initial shift of the YSR states deeper into the superconducting energy gap. However, the YSR states do not reach or cross zero energy. The YSR states come closest to zero energy for $\Delta z=-150$~pm before shifting back to higher energies as a result of the molecule being pushed back to the surface (Fig.\ref{fig:1}b). An analysis of the asymmetry confirms that the system does not pass through the quantum phase transition (\Fig{3}c). Indeed, the asymmetry does not change sign at $\Delta z=-150$~pm and the hole-like excitation remains stronger than the electron-like excitation throughout (see detailed spectra in \Fig{3}d). A change in asymmetry occurs only upon further approach (\Fig{3}c,d) when resonant Andreev reflections become the dominant tunneling process~\cite{Ruby2015}.

Importantly, the overall trend of the YSR shift resembles the case with the tip above the Fe center, reflecting an analogous dependence on the tip--molecule force. Hence, despite the inverse asymmetry, the YSR state reflects the same ground state. Given that we observe the same energy on ligand and center, we suggest that the YSR state arises from a delocalized spin state associated with Fe $d_\pi$ states hybridized with $\pi$ states of the ligand~\cite{Ali2012,Bhandary2016,Rubio-Verdu2018}. The absence of a quantum phase transition on the ligand reflects differences in the elastic response of the molecule to the applied force.  

 \begin{figure}[tb]
	\includegraphics[width=\columnwidth]{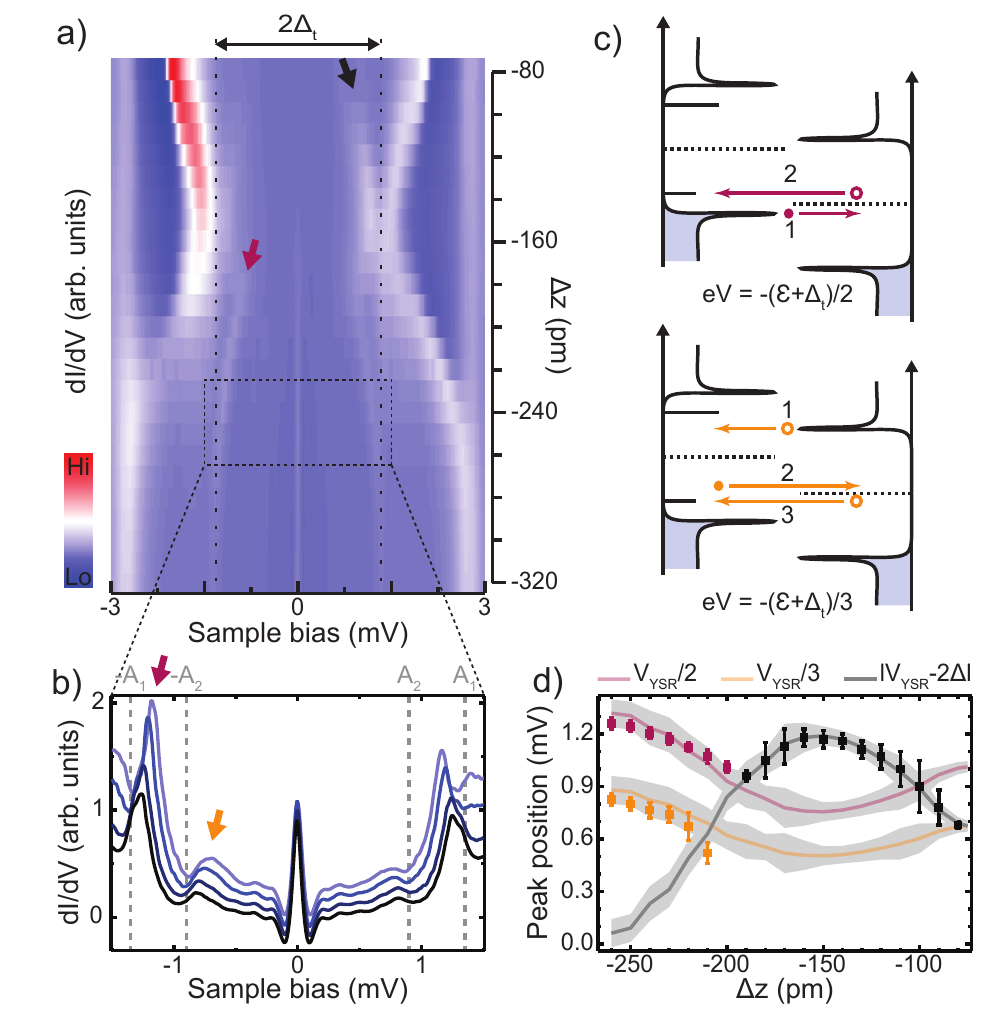}
	\caption{a) Closer view of the approach set in \Fig{3}b. b) Four spectra (offset for clarity) of this set showing two peaks due to the Andreev processes depicted in c). An electron (hole) is reflected through the junction until its energy matches the YSR state. d) Extracted positions of the peaks shown in a and b (rectangles) as well as traces showing the expected positions of a thermal excitation of the YSR state (black trace) and of the processes shown in c (yellow and pink) given the position of the YSR state.}
	\label{fig:4}
\end{figure}

So far, we could understand all spectral features in terms of single-electron excitations of the YSR states. We already noted that on both Fe center and ligand, the asymmetry is reduced at tip approaches of $\Delta z \approx-190\upm$. The continuous change of spectral intensity can be explained by resonant Andreev reflections gaining strength as the tunnel conductance increases~\cite{Ruby2015}. Furthermore, we observe the onset of a zero-bias resonance as a fingerprint of Cooper pair (Josephson) tunneling upon further increase of the junction conductance. (Refs.\ \cite{Randeria2016,Brand2018} also observed Josephson peaks in STM spectra of single adatoms and molecules at close tip--sample distance.) In addition, we find peaks at $eV=\pm\DeltaT$. We interpret these as the threshold for the lowest-order ($n=2$) multiple Andreev reflections (MAR), generally expected at $eV_n=\pm (\Delta+\DeltaT)/{n}$ for an $(n-1)$-fold Andreev reflection ($n=2,3,...$) \cite{Ternes2006}. We observe additional subgap peaks, which shift in energy with tip approach. The set of peaks following $eV=\DeltaT-\epsilon$ arises from thermal excitations~\cite{Ruby2015}. Interestingly, the contrast-enhanced  plot of \didv spectra (\Fig{4}a,b) shows two other sets of peaks, which follow the relation $eV=({\DeltaT+\epsilon})/{n}$ with $n=2,3$~\cite{Randeria2016} (\Fig{4}d). 
They originate from the excitation of the YSR state by electron/holes that are Andreev reflected from the superconductor (see sketches in \Fig{4}c). Unlike conventional MARs, these processes reflect the asymmetry of the YSR states and might also be usable as fingerprints of the quantum phase transition.

Finally, when the tip is in contact with the molecule (as deduced from the flat $G-\Delta z$ curves in \Fig{2}a and \Fig{3}a), the Andreev reflections through the YSR states merge with the conventional MARs. In this contact regime, the coupling of the impurity to the tip and the surface start to compete which each another, leading to an effective coupling of the two electrodes. As a result, an applied bias drives the junction out of equilibrium so that signatures of local excitations can be no longer detected in spectroscopic measurements ~\cite{Andersen2011}.

In conclusion, we have realized a magnetic-impurity junction whose exchange coupling can be precisely, continuously, and reversibly tuned through mechanical forces exerted by an STM tip. Moreover, the junction conductance can be varied by almost three orders of magnitude which enables us to access various transport regimes. We have employed this setup for a detailed investigation of the quantum phase transition between the screened-spin and free-spin ground states. We confirm the first-order nature of the transition in that the spectral weights of the single-particle addition spectrum change abruptly at the transition. We do not observe the predicted discontinuous jumps in the YSR energy and the local order parameter. Our analytical estimates show that the expected jumps in YSR energies are by orders of magnitude smaller than the energy resolution even at mK temperatures. Finally, we emphasize that our results highlight a method for unambiguously determining the nature of the ground state through variation of the exchange coupling.

\acknowledgements
We thank S.\ Loth for discussions. Financial support by Deutsche Forschungsgemeinschaft through grants HE7368/2 and CRC 183, as well as by the European Research Council through the consolidator grant ``NanoSpin" is gratefully acknowledged. FvO performed part of this work at the Aspen Center for Physics supported by NSF grant PHY-1607611.

\bibliographystyle{apsrev4-1}


\begin{thebibliography}{43}%
\makeatletter
\providecommand \@ifxundefined [1]{%
 \@ifx{#1\undefined}
}%
\providecommand \@ifnum [1]{%
 \ifnum #1\expandafter \@firstoftwo
 \else \expandafter \@secondoftwo
 \fi
}%
\providecommand \@ifx [1]{%
 \ifx #1\expandafter \@firstoftwo
 \else \expandafter \@secondoftwo
 \fi
}%
\providecommand \natexlab [1]{#1}%
\providecommand \enquote  [1]{``#1''}%
\providecommand \bibnamefont  [1]{#1}%
\providecommand \bibfnamefont [1]{#1}%
\providecommand \citenamefont [1]{#1}%
\providecommand \href@noop [0]{\@secondoftwo}%
\providecommand \href [0]{\begingroup \@sanitize@url \@href}%
\providecommand \@href[1]{\@@startlink{#1}\@@href}%
\providecommand \@@href[1]{\endgroup#1\@@endlink}%
\providecommand \@sanitize@url [0]{\catcode `\\12\catcode `\$12\catcode
  `\&12\catcode `\#12\catcode `\^12\catcode `\_12\catcode `\%12\relax}%
\providecommand \@@startlink[1]{}%
\providecommand \@@endlink[0]{}%
\providecommand \url  [0]{\begingroup\@sanitize@url \@url }%
\providecommand \@url [1]{\endgroup\@href {#1}{\urlprefix }}%
\providecommand \urlprefix  [0]{URL }%
\providecommand \Eprint [0]{\href }%
\providecommand \doibase [0]{http://dx.doi.org/}%
\providecommand \selectlanguage [0]{\@gobble}%
\providecommand \bibinfo  [0]{\@secondoftwo}%
\providecommand \bibfield  [0]{\@secondoftwo}%
\providecommand \translation [1]{[#1]}%
\providecommand \BibitemOpen [0]{}%
\providecommand \bibitemStop [0]{}%
\providecommand \bibitemNoStop [0]{.\EOS\space}%
\providecommand \EOS [0]{\spacefactor3000\relax}%
\providecommand \BibitemShut  [1]{\csname bibitem#1\endcsname}%
\let\auto@bib@innerbib\@empty
\bibitem [{\citenamefont {Yu}(1965)}]{Yu1965}%
  \BibitemOpen
  \bibfield  {author} {\bibinfo {author} {\bibfnamefont {L.}~\bibnamefont
  {Yu}},\ }\href {\doibase 10.7498/aps.21.75} {\bibfield  {journal} {\bibinfo
  {journal} {Acta Phys. Sin.}\ }\textbf {\bibinfo {volume} {21}},\ \bibinfo
  {pages} {75} (\bibinfo {year} {1965})}\BibitemShut {NoStop}%
\bibitem [{\citenamefont {Shiba}(1968)}]{Shiba1968}%
  \BibitemOpen
  \bibfield  {author} {\bibinfo {author} {\bibfnamefont {H.}~\bibnamefont
  {Shiba}},\ }\href {\doibase 10.1143/PTP.40.435} {\bibfield  {journal}
  {\bibinfo  {journal} {Prog. Theor. Phys.}\ }\textbf {\bibinfo {volume}
  {40}},\ \bibinfo {pages} {435} (\bibinfo {year} {1968})}\BibitemShut
  {NoStop}%
\bibitem [{\citenamefont {Rusinov}(1968)}]{Rusinov1969}%
  \BibitemOpen
  \bibfield  {author} {\bibinfo {author} {\bibfnamefont {A.~I.}\ \bibnamefont
  {Rusinov}},\ }\href@noop {} {\bibfield  {journal} {\bibinfo  {journal} {Sov.
  J. Exp. Theor. Phys.}\ }\textbf {\bibinfo {volume} {9}},\ \bibinfo {pages}
  {85} (\bibinfo {year} {1969})}\BibitemShut {NoStop}%
\bibitem [{\citenamefont {Sakurai}(1970)}]{Sakurai1970}%
  \BibitemOpen
  \bibfield  {author} {\bibinfo {author} {\bibfnamefont {A.}~\bibnamefont
  {Sakurai}},\ }\href {\doibase 10.1143/PTP.44.1472} {\bibfield  {journal}
  {\bibinfo  {journal} {Prog. Theor. Phys.}\ }\textbf {\bibinfo {volume}
  {44}},\ \bibinfo {pages} {1472} (\bibinfo {year} {1970})}\BibitemShut
  {NoStop}%
\bibitem [{\citenamefont {Matsuura}(1977)}]{Matsuura1977}%
  \BibitemOpen
  \bibfield  {author} {\bibinfo {author} {\bibfnamefont {T.}~\bibnamefont
  {Matsuura}},\ }\href {\doibase 10.1143/PTP.57.1823} {\bibfield  {journal}
  {\bibinfo  {journal} {Prog. Theor. Phys.}\ }\textbf {\bibinfo {volume}
  {57}},\ \bibinfo {pages} {1823} (\bibinfo {year} {1977})}\BibitemShut
  {NoStop}%
\bibitem [{\citenamefont {Sakai}\ \emph {et~al.}(1993)\citenamefont {Sakai},
  \citenamefont {Shimizu}, \citenamefont {Shiba},\ and\ \citenamefont
  {Satori}}]{Sakai1993}%
  \BibitemOpen
  \bibfield  {author} {\bibinfo {author} {\bibfnamefont {O.}~\bibnamefont
  {Sakai}}, \bibinfo {author} {\bibfnamefont {Y.}~\bibnamefont {Shimizu}},
  \bibinfo {author} {\bibfnamefont {H.}~\bibnamefont {Shiba}}, \ and\ \bibinfo
  {author} {\bibfnamefont {K.}~\bibnamefont {Satori}},\ }\href {\doibase
  10.1143/JPSJ.62.3181} {\bibfield  {journal} {\bibinfo  {journal} {J. Phys.
  Soc. Jpn.}\ }\textbf {\bibinfo {volume} {62}},\ \bibinfo {pages} {3181}
  (\bibinfo {year} {1993})}\BibitemShut {NoStop}%
\bibitem [{\citenamefont {Balatsky}\ \emph {et~al.}(2006)\citenamefont
  {Balatsky}, \citenamefont {Vekhter},\ and\ \citenamefont
  {Zhu}}]{Balatsky2006}%
  \BibitemOpen
  \bibfield  {author} {\bibinfo {author} {\bibfnamefont {A.~V.}\ \bibnamefont
  {Balatsky}}, \bibinfo {author} {\bibfnamefont {I.}~\bibnamefont {Vekhter}}, \
  and\ \bibinfo {author} {\bibfnamefont {J.~X.}\ \bibnamefont {Zhu}},\ }\href
  {\doibase 10.1103/RevModPhys.78.373} {\bibfield  {journal} {\bibinfo
  {journal} {Rev. Mod. Phys.}\ }\textbf {\bibinfo {volume} {78}},\ \bibinfo
  {pages} {373} (\bibinfo {year} {2006})}\BibitemShut {NoStop}%
\bibitem [{\citenamefont {Heinrich}\ \emph {et~al.}(2018)\citenamefont
  {Heinrich}, \citenamefont {Pascual},\ and\ \citenamefont
  {Franke}}]{Heinrich2018}%
  \BibitemOpen
  \bibfield  {author} {\bibinfo {author} {\bibfnamefont {B.~W.}\ \bibnamefont
  {Heinrich}}, \bibinfo {author} {\bibfnamefont {J.~I.}\ \bibnamefont
  {Pascual}}, \ and\ \bibinfo {author} {\bibfnamefont {K.~J.}\ \bibnamefont
  {Franke}},\ }\href {\doibase 10.1016/j.progsurf.2018.01.001} {\bibfield
  {journal} {\bibinfo  {journal} {Prog. Surf. Sci.}\ }\textbf {\bibinfo
  {volume} {93}},\ \bibinfo {pages} {1} (\bibinfo {year} {2018})}\BibitemShut
  {NoStop}%
\bibitem [{\citenamefont {Salkola}\ \emph {et~al.}(1997)\citenamefont
  {Salkola}, \citenamefont {Balatsky},\ and\ \citenamefont
  {Schrieffer}}]{Salkola1997}%
  \BibitemOpen
  \bibfield  {author} {\bibinfo {author} {\bibfnamefont {M.~I.}\ \bibnamefont
  {Salkola}}, \bibinfo {author} {\bibfnamefont {A.~V.}\ \bibnamefont
  {Balatsky}}, \ and\ \bibinfo {author} {\bibfnamefont {J.~R.}\ \bibnamefont
  {Schrieffer}},\ }\href {\doibase 10.1103/PhysRevB.55.12648} {\bibfield
  {journal} {\bibinfo  {journal} {Phys. Rev. B}\ }\textbf {\bibinfo {volume}
  {55}},\ \bibinfo {pages} {12648} (\bibinfo {year} {1997})}\BibitemShut
  {NoStop}%
\bibitem [{\citenamefont {Flatt{\'{e}}}\ and\ \citenamefont
  {Byers}(1997)}]{Flatte1997}%
  \BibitemOpen
  \bibfield  {author} {\bibinfo {author} {\bibfnamefont {M.~E.}\ \bibnamefont
  {Flatt{\'{e}}}}\ and\ \bibinfo {author} {\bibfnamefont {J.~M.}\ \bibnamefont
  {Byers}},\ }\href {\doibase https://doi.org/10.1103/PhysRevB.56.11213}
  {\bibfield  {journal} {\bibinfo  {journal} {Phys. Rev. B}\ }\textbf {\bibinfo
  {volume} {56}},\ \bibinfo {pages} {11213} (\bibinfo {year}
  {1997})}\BibitemShut {NoStop}%
\bibitem [{\citenamefont {Meng}\ \emph {et~al.}(2015)\citenamefont {Meng},
  \citenamefont {Klinovaja}, \citenamefont {Hoffman}, \citenamefont {Simon},\
  and\ \citenamefont {Loss}}]{Meng2015}%
  \BibitemOpen
  \bibfield  {author} {\bibinfo {author} {\bibfnamefont {T.}~\bibnamefont
  {Meng}}, \bibinfo {author} {\bibfnamefont {J.}~\bibnamefont {Klinovaja}},
  \bibinfo {author} {\bibfnamefont {S.}~\bibnamefont {Hoffman}}, \bibinfo
  {author} {\bibfnamefont {P.}~\bibnamefont {Simon}}, \ and\ \bibinfo {author}
  {\bibfnamefont {D.}~\bibnamefont {Loss}},\ }\href {\doibase
  10.1103/PhysRevB.92.064503} {\bibfield  {journal} {\bibinfo  {journal} {Phys.
  Rev. B}\ }\textbf {\bibinfo {volume} {92}},\ \bibinfo {pages} {064503}
  (\bibinfo {year} {2015})}\BibitemShut {NoStop}%
\bibitem [{\citenamefont {Maurand}\ \emph {et~al.}(2012)\citenamefont
  {Maurand}, \citenamefont {Meng}, \citenamefont {Bonet}, \citenamefont
  {Florens}, \citenamefont {Marty},\ and\ \citenamefont
  {Wernsdorfer}}]{Maurand2012}%
  \BibitemOpen
  \bibfield  {author} {\bibinfo {author} {\bibfnamefont {R.}~\bibnamefont
  {Maurand}}, \bibinfo {author} {\bibfnamefont {T.}~\bibnamefont {Meng}},
  \bibinfo {author} {\bibfnamefont {E.}~\bibnamefont {Bonet}}, \bibinfo
  {author} {\bibfnamefont {S.}~\bibnamefont {Florens}}, \bibinfo {author}
  {\bibfnamefont {L.}~\bibnamefont {Marty}}, \ and\ \bibinfo {author}
  {\bibfnamefont {W.}~\bibnamefont {Wernsdorfer}},\ }\href {\doibase
  10.1103/PhysRevX.2.011009} {\bibfield  {journal} {\bibinfo  {journal} {Phys.
  Rev. X}\ }\textbf {\bibinfo {volume} {2}},\ \bibinfo {pages} {011009}
  (\bibinfo {year} {2012})}\BibitemShut {NoStop}%
\bibitem [{\citenamefont {Delagrange}\ \emph {et~al.}(2015)\citenamefont
  {Delagrange}, \citenamefont {Luitz}, \citenamefont {Weil}, \citenamefont
  {Kasumov}, \citenamefont {Meden}, \citenamefont {Bouchiat},\ and\
  \citenamefont {Deblock}}]{Delagrange2015}%
  \BibitemOpen
  \bibfield  {author} {\bibinfo {author} {\bibfnamefont {R.}~\bibnamefont
  {Delagrange}}, \bibinfo {author} {\bibfnamefont {D.~J.}\ \bibnamefont
  {Luitz}}, \bibinfo {author} {\bibfnamefont {R.}~\bibnamefont {Weil}},
  \bibinfo {author} {\bibfnamefont {A.}~\bibnamefont {Kasumov}}, \bibinfo
  {author} {\bibfnamefont {V.}~\bibnamefont {Meden}}, \bibinfo {author}
  {\bibfnamefont {H.}~\bibnamefont {Bouchiat}}, \ and\ \bibinfo {author}
  {\bibfnamefont {R.}~\bibnamefont {Deblock}},\ }\href {\doibase
  10.1103/PhysRevB.91.241401} {\bibfield  {journal} {\bibinfo  {journal} {Phys.
  Rev. B}\ }\textbf {\bibinfo {volume} {91}},\ \bibinfo {pages} {241401}
  (\bibinfo {year} {2015})}\BibitemShut {NoStop}%
\bibitem [{\citenamefont {Delagrange}\ \emph {et~al.}(2016)\citenamefont
  {Delagrange}, \citenamefont {Weil}, \citenamefont {Kasumov}, \citenamefont
  {Ferrier}, \citenamefont {Bouchiat},\ and\ \citenamefont
  {Deblock}}]{Delagrange2016}%
  \BibitemOpen
  \bibfield  {author} {\bibinfo {author} {\bibfnamefont {R.}~\bibnamefont
  {Delagrange}}, \bibinfo {author} {\bibfnamefont {R.}~\bibnamefont {Weil}},
  \bibinfo {author} {\bibfnamefont {A.}~\bibnamefont {Kasumov}}, \bibinfo
  {author} {\bibfnamefont {M.}~\bibnamefont {Ferrier}}, \bibinfo {author}
  {\bibfnamefont {H.}~\bibnamefont {Bouchiat}}, \ and\ \bibinfo {author}
  {\bibfnamefont {R.}~\bibnamefont {Deblock}},\ }\href {\doibase
  10.1103/PhysRevB.93.195437} {\bibfield  {journal} {\bibinfo  {journal} {Phys.
  Rev. B}\ }\textbf {\bibinfo {volume} {93}},\ \bibinfo {pages} {195437}
  (\bibinfo {year} {2016})}\BibitemShut {NoStop}%
\bibitem [{\citenamefont {Deacon}\ \emph {et~al.}(2010)\citenamefont {Deacon},
  \citenamefont {Tanaka}, \citenamefont {Oiwa}, \citenamefont {Sakano},
  \citenamefont {Yoshida}, \citenamefont {Shibata}, \citenamefont {Hirakawa},\
  and\ \citenamefont {Tarucha}}]{Deacon2010}%
  \BibitemOpen
  \bibfield  {author} {\bibinfo {author} {\bibfnamefont {R.~S.}\ \bibnamefont
  {Deacon}}, \bibinfo {author} {\bibfnamefont {Y.}~\bibnamefont {Tanaka}},
  \bibinfo {author} {\bibfnamefont {A.}~\bibnamefont {Oiwa}}, \bibinfo {author}
  {\bibfnamefont {R.}~\bibnamefont {Sakano}}, \bibinfo {author} {\bibfnamefont
  {K.}~\bibnamefont {Yoshida}}, \bibinfo {author} {\bibfnamefont
  {K.}~\bibnamefont {Shibata}}, \bibinfo {author} {\bibfnamefont
  {K.}~\bibnamefont {Hirakawa}}, \ and\ \bibinfo {author} {\bibfnamefont
  {S.}~\bibnamefont {Tarucha}},\ }\href {\doibase
  10.1103/PhysRevLett.104.076805} {\bibfield  {journal} {\bibinfo  {journal}
  {Phys. Rev. Lett.}\ }\textbf {\bibinfo {volume} {104}},\ \bibinfo {pages}
  {076805} (\bibinfo {year} {2010})}\BibitemShut {NoStop}%
\bibitem [{\citenamefont {Lee}\ \emph {et~al.}(2014)\citenamefont {Lee},
  \citenamefont {Jiang}, \citenamefont {Houzet}, \citenamefont {Aguado},
  \citenamefont {Lieber},\ and\ \citenamefont {{De Franceschi}}}]{Lee2014}%
  \BibitemOpen
  \bibfield  {author} {\bibinfo {author} {\bibfnamefont {E.~J.}\ \bibnamefont
  {Lee}}, \bibinfo {author} {\bibfnamefont {X.}~\bibnamefont {Jiang}}, \bibinfo
  {author} {\bibfnamefont {M.}~\bibnamefont {Houzet}}, \bibinfo {author}
  {\bibfnamefont {R.}~\bibnamefont {Aguado}}, \bibinfo {author} {\bibfnamefont
  {C.~M.}\ \bibnamefont {Lieber}}, \ and\ \bibinfo {author} {\bibfnamefont
  {S.}~\bibnamefont {{De Franceschi}}},\ }\href {\doibase
  10.1038/nnano.2013.267} {\bibfield  {journal} {\bibinfo  {journal} {Nat.
  Nanotechnol.}\ }\textbf {\bibinfo {volume} {9}},\ \bibinfo {pages} {79}
  (\bibinfo {year} {2014})}\BibitemShut {NoStop}%
\bibitem [{\citenamefont {Lee}\ \emph {et~al.}(2017)\citenamefont {Lee},
  \citenamefont {Jiang}, \citenamefont {\ifmmode~\check{Z}\else
  \v{Z}\fi{}itko}, \citenamefont {Aguado}, \citenamefont {Lieber},\ and\
  \citenamefont {De~Franceschi}}]{Lee2017}%
  \BibitemOpen
  \bibfield  {author} {\bibinfo {author} {\bibfnamefont {E.~J.~H.}\
  \bibnamefont {Lee}}, \bibinfo {author} {\bibfnamefont {X.}~\bibnamefont
  {Jiang}}, \bibinfo {author} {\bibfnamefont {R.}~\bibnamefont
  {\ifmmode~\check{Z}\else \v{Z}\fi{}itko}}, \bibinfo {author} {\bibfnamefont
  {R.}~\bibnamefont {Aguado}}, \bibinfo {author} {\bibfnamefont {C.~M.}\
  \bibnamefont {Lieber}}, \ and\ \bibinfo {author} {\bibfnamefont
  {S.}~\bibnamefont {De~Franceschi}},\ }\href {\doibase
  10.1103/PhysRevB.95.180502} {\bibfield  {journal} {\bibinfo  {journal} {Phys.
  Rev. B}\ }\textbf {\bibinfo {volume} {95}},\ \bibinfo {pages} {180502}
  (\bibinfo {year} {2017})}\BibitemShut {NoStop}%
\bibitem [{\citenamefont {Ruby}\ \emph
  {et~al.}(2015{\natexlab{a}})\citenamefont {Ruby}, \citenamefont {Pientka},
  \citenamefont {Peng}, \citenamefont {von Oppen}, \citenamefont {Heinrich},\
  and\ \citenamefont {Franke}}]{Ruby2015}%
  \BibitemOpen
  \bibfield  {author} {\bibinfo {author} {\bibfnamefont {M.}~\bibnamefont
  {Ruby}}, \bibinfo {author} {\bibfnamefont {F.}~\bibnamefont {Pientka}},
  \bibinfo {author} {\bibfnamefont {Y.}~\bibnamefont {Peng}}, \bibinfo {author}
  {\bibfnamefont {F.}~\bibnamefont {von Oppen}}, \bibinfo {author}
  {\bibfnamefont {B.~W.}\ \bibnamefont {Heinrich}}, \ and\ \bibinfo {author}
  {\bibfnamefont {K.~J.}\ \bibnamefont {Franke}},\ }\href {\doibase
  10.1103/PhysRevLett.115.087001} {\bibfield  {journal} {\bibinfo  {journal}
  {Phys. Rev. Lett.}\ }\textbf {\bibinfo {volume} {115}},\ \bibinfo {pages}
  {087001} (\bibinfo {year} {2015}{\natexlab{a}})}\BibitemShut {NoStop}%
\bibitem [{\citenamefont {Yazdani}\ \emph {et~al.}(1997)\citenamefont
  {Yazdani}, \citenamefont {Jones}, \citenamefont {Lutz}, \citenamefont
  {Crommie},\ and\ \citenamefont {Eigler}}]{Yazdani1997}%
  \BibitemOpen
  \bibfield  {author} {\bibinfo {author} {\bibfnamefont {A.}~\bibnamefont
  {Yazdani}}, \bibinfo {author} {\bibfnamefont {B.~A.}\ \bibnamefont {Jones}},
  \bibinfo {author} {\bibfnamefont {C.~P.}\ \bibnamefont {Lutz}}, \bibinfo
  {author} {\bibfnamefont {M.~F.}\ \bibnamefont {Crommie}}, \ and\ \bibinfo
  {author} {\bibfnamefont {D.~M.}\ \bibnamefont {Eigler}},\ }\href {\doibase
  10.1126/science.275.5307.1767} {\bibfield  {journal} {\bibinfo  {journal}
  {Science}\ }\textbf {\bibinfo {volume} {275}},\ \bibinfo {pages} {1767}
  (\bibinfo {year} {1997})}\BibitemShut {NoStop}%
\bibitem [{\citenamefont {Ji}\ \emph {et~al.}(2008)\citenamefont {Ji},
  \citenamefont {Zhang}, \citenamefont {Fu}, \citenamefont {Chen},
  \citenamefont {Ma}, \citenamefont {Li}, \citenamefont {Duan}, \citenamefont
  {Jia},\ and\ \citenamefont {Xue}}]{Ji2008}%
  \BibitemOpen
  \bibfield  {author} {\bibinfo {author} {\bibfnamefont {S.-H.}\ \bibnamefont
  {Ji}}, \bibinfo {author} {\bibfnamefont {T.}~\bibnamefont {Zhang}}, \bibinfo
  {author} {\bibfnamefont {Y.-S.}\ \bibnamefont {Fu}}, \bibinfo {author}
  {\bibfnamefont {X.}~\bibnamefont {Chen}}, \bibinfo {author} {\bibfnamefont
  {X.-C.}\ \bibnamefont {Ma}}, \bibinfo {author} {\bibfnamefont
  {J.}~\bibnamefont {Li}}, \bibinfo {author} {\bibfnamefont {W.-H.}\
  \bibnamefont {Duan}}, \bibinfo {author} {\bibfnamefont {J.-F.}\ \bibnamefont
  {Jia}}, \ and\ \bibinfo {author} {\bibfnamefont {Q.-K.}\ \bibnamefont
  {Xue}},\ }\href {\doibase 10.1103/PhysRevLett.100.226801} {\bibfield
  {journal} {\bibinfo  {journal} {Phys. Rev. Lett.}\ }\textbf {\bibinfo
  {volume} {100}},\ \bibinfo {pages} {226801} (\bibinfo {year}
  {2008})}\BibitemShut {NoStop}%
\bibitem [{\citenamefont {Franke}\ \emph {et~al.}(2011)\citenamefont {Franke},
  \citenamefont {Schulze},\ and\ \citenamefont {Pascual}}]{Franke2011}%
  \BibitemOpen
  \bibfield  {author} {\bibinfo {author} {\bibfnamefont {K.~J.}\ \bibnamefont
  {Franke}}, \bibinfo {author} {\bibfnamefont {G.}~\bibnamefont {Schulze}}, \
  and\ \bibinfo {author} {\bibfnamefont {J.~I.}\ \bibnamefont {Pascual}},\
  }\href {\doibase 10.1126/science.1202204} {\bibfield  {journal} {\bibinfo
  {journal} {Science}\ }\textbf {\bibinfo {volume} {332}},\ \bibinfo {pages}
  {940} (\bibinfo {year} {2011})}\BibitemShut {NoStop}%
\bibitem [{\citenamefont {M{\'{e}}nard}\ \emph {et~al.}(2015)\citenamefont
  {M{\'{e}}nard}, \citenamefont {Guissart}, \citenamefont {Brun}, \citenamefont
  {Pons}, \citenamefont {Stolyarov}, \citenamefont {Debontridder},
  \citenamefont {Leclerc}, \citenamefont {Janod}, \citenamefont {Cario},
  \citenamefont {Roditchev}, \citenamefont {Simon},\ and\ \citenamefont
  {Cren}}]{Menard2015}%
  \BibitemOpen
  \bibfield  {author} {\bibinfo {author} {\bibfnamefont {G.~C.}\ \bibnamefont
  {M{\'{e}}nard}}, \bibinfo {author} {\bibfnamefont {S.}~\bibnamefont
  {Guissart}}, \bibinfo {author} {\bibfnamefont {C.}~\bibnamefont {Brun}},
  \bibinfo {author} {\bibfnamefont {S.}~\bibnamefont {Pons}}, \bibinfo {author}
  {\bibfnamefont {V.~S.}\ \bibnamefont {Stolyarov}}, \bibinfo {author}
  {\bibfnamefont {F.}~\bibnamefont {Debontridder}}, \bibinfo {author}
  {\bibfnamefont {M.~V.}\ \bibnamefont {Leclerc}}, \bibinfo {author}
  {\bibfnamefont {E.}~\bibnamefont {Janod}}, \bibinfo {author} {\bibfnamefont
  {L.}~\bibnamefont {Cario}}, \bibinfo {author} {\bibfnamefont
  {D.}~\bibnamefont {Roditchev}}, \bibinfo {author} {\bibfnamefont
  {P.}~\bibnamefont {Simon}}, \ and\ \bibinfo {author} {\bibfnamefont
  {T.}~\bibnamefont {Cren}},\ }\href {\doibase 10.1038/nphys3508} {\bibfield
  {journal} {\bibinfo  {journal} {Nat. Phys.}\ }\textbf {\bibinfo {volume}
  {11}},\ \bibinfo {pages} {1013} (\bibinfo {year} {2015})}\BibitemShut
  {NoStop}%
\bibitem [{\citenamefont {Ruby}\ \emph {et~al.}(2016)\citenamefont {Ruby},
  \citenamefont {Peng}, \citenamefont {von Oppen}, \citenamefont {Heinrich},\
  and\ \citenamefont {Franke}}]{Ruby2016}%
  \BibitemOpen
  \bibfield  {author} {\bibinfo {author} {\bibfnamefont {M.}~\bibnamefont
  {Ruby}}, \bibinfo {author} {\bibfnamefont {Y.}~\bibnamefont {Peng}}, \bibinfo
  {author} {\bibfnamefont {F.}~\bibnamefont {von Oppen}}, \bibinfo {author}
  {\bibfnamefont {B.~W.}\ \bibnamefont {Heinrich}}, \ and\ \bibinfo {author}
  {\bibfnamefont {K.~J.}\ \bibnamefont {Franke}},\ }\href {\doibase
  10.1103/PhysRevLett.117.186801} {\bibfield  {journal} {\bibinfo  {journal}
  {Phys. Rev. Lett.}\ }\textbf {\bibinfo {volume} {117}},\ \bibinfo {pages}
  {186801} (\bibinfo {year} {2016})}\BibitemShut {NoStop}%
\bibitem [{\citenamefont {Choi}\ \emph {et~al.}(2017)\citenamefont {Choi},
  \citenamefont {Rubio-Verd{\'{u}}}, \citenamefont {{De Bruijckere}},
  \citenamefont {Ugeda}, \citenamefont {Lorente},\ and\ \citenamefont
  {Pascual}}]{Choi2017}%
  \BibitemOpen
  \bibfield  {author} {\bibinfo {author} {\bibfnamefont {D.~J.}\ \bibnamefont
  {Choi}}, \bibinfo {author} {\bibfnamefont {C.}~\bibnamefont
  {Rubio-Verd{\'{u}}}}, \bibinfo {author} {\bibfnamefont {J.}~\bibnamefont {{De
  Bruijckere}}}, \bibinfo {author} {\bibfnamefont {M.~M.}\ \bibnamefont
  {Ugeda}}, \bibinfo {author} {\bibfnamefont {N.}~\bibnamefont {Lorente}}, \
  and\ \bibinfo {author} {\bibfnamefont {J.~I.}\ \bibnamefont {Pascual}},\
  }\href {\doibase 10.1038/ncomms15175} {\bibfield  {journal} {\bibinfo
  {journal} {Nat. Commun.}\ }\textbf {\bibinfo {volume} {8}},\ \bibinfo {pages}
  {15175} (\bibinfo {year} {2017})}\BibitemShut {NoStop}%
\bibitem [{\citenamefont {Cornils}\ \emph {et~al.}(2017)\citenamefont
  {Cornils}, \citenamefont {Kamlapure}, \citenamefont {Zhou}, \citenamefont
  {Pradhan}, \citenamefont {Khajetoorians}, \citenamefont {Fransson},
  \citenamefont {Wiebe},\ and\ \citenamefont {Wiesendanger}}]{Cornils2017}%
  \BibitemOpen
  \bibfield  {author} {\bibinfo {author} {\bibfnamefont {L.}~\bibnamefont
  {Cornils}}, \bibinfo {author} {\bibfnamefont {A.}~\bibnamefont {Kamlapure}},
  \bibinfo {author} {\bibfnamefont {L.}~\bibnamefont {Zhou}}, \bibinfo {author}
  {\bibfnamefont {S.}~\bibnamefont {Pradhan}}, \bibinfo {author} {\bibfnamefont
  {A.~A.}\ \bibnamefont {Khajetoorians}}, \bibinfo {author} {\bibfnamefont
  {J.}~\bibnamefont {Fransson}}, \bibinfo {author} {\bibfnamefont
  {J.}~\bibnamefont {Wiebe}}, \ and\ \bibinfo {author} {\bibfnamefont
  {R.}~\bibnamefont {Wiesendanger}},\ }\href {\doibase
  10.1103/PhysRevLett.119.197002} {\bibfield  {journal} {\bibinfo  {journal}
  {Phys. Rev. Lett.}\ }\textbf {\bibinfo {volume} {119}},\ \bibinfo {pages}
  {197002} (\bibinfo {year} {2017})}\BibitemShut {NoStop}%
\bibitem [{\citenamefont {Ruby}\ \emph
  {et~al.}(2015{\natexlab{b}})\citenamefont {Ruby}, \citenamefont {Pientka},
  \citenamefont {Peng}, \citenamefont {von Oppen}, \citenamefont {Heinrich},\
  and\ \citenamefont {Franke}}]{Ruby2015chains}%
  \BibitemOpen
  \bibfield  {author} {\bibinfo {author} {\bibfnamefont {M.}~\bibnamefont
  {Ruby}}, \bibinfo {author} {\bibfnamefont {F.}~\bibnamefont {Pientka}},
  \bibinfo {author} {\bibfnamefont {Y.}~\bibnamefont {Peng}}, \bibinfo {author}
  {\bibfnamefont {F.}~\bibnamefont {von Oppen}}, \bibinfo {author}
  {\bibfnamefont {B.~W.}\ \bibnamefont {Heinrich}}, \ and\ \bibinfo {author}
  {\bibfnamefont {K.~J.}\ \bibnamefont {Franke}},\ }\href {\doibase
  10.1103/PhysRevLett.115.197204} {\bibfield  {journal} {\bibinfo  {journal}
  {Phys. Rev. Lett.}\ }\textbf {\bibinfo {volume} {115}},\ \bibinfo {pages}
  {197204} (\bibinfo {year} {2015}{\natexlab{b}})}\BibitemShut {NoStop}%
\bibitem [{\citenamefont {Hatter}\ \emph {et~al.}(2015)\citenamefont {Hatter},
  \citenamefont {Heinrich}, \citenamefont {Ruby}, \citenamefont {Pascual},\
  and\ \citenamefont {Franke}}]{Hatter2015}%
  \BibitemOpen
  \bibfield  {author} {\bibinfo {author} {\bibfnamefont {N.}~\bibnamefont
  {Hatter}}, \bibinfo {author} {\bibfnamefont {B.~W.}\ \bibnamefont
  {Heinrich}}, \bibinfo {author} {\bibfnamefont {M.}~\bibnamefont {Ruby}},
  \bibinfo {author} {\bibfnamefont {J.~I.}\ \bibnamefont {Pascual}}, \ and\
  \bibinfo {author} {\bibfnamefont {K.~J.}\ \bibnamefont {Franke}},\ }\href
  {\doibase 10.1038/ncomms9988} {\bibfield  {journal} {\bibinfo  {journal}
  {Nat. Commun.}\ }\textbf {\bibinfo {volume} {6}},\ \bibinfo {pages} {8988}
  (\bibinfo {year} {2015})}\BibitemShut {NoStop}%
\bibitem [{\citenamefont {Hatter}\ \emph {et~al.}(2017)\citenamefont {Hatter},
  \citenamefont {Heinrich}, \citenamefont {Rolf},\ and\ \citenamefont
  {Franke}}]{Hatter2017}%
  \BibitemOpen
  \bibfield  {author} {\bibinfo {author} {\bibfnamefont {N.}~\bibnamefont
  {Hatter}}, \bibinfo {author} {\bibfnamefont {B.~W.}\ \bibnamefont
  {Heinrich}}, \bibinfo {author} {\bibfnamefont {D.}~\bibnamefont {Rolf}}, \
  and\ \bibinfo {author} {\bibfnamefont {K.~J.}\ \bibnamefont {Franke}},\
  }\href {\doibase 10.1038/s41467-017-02277-7} {\bibfield  {journal} {\bibinfo
  {journal} {Nat. Commun.}\ }\textbf {\bibinfo {volume} {8}},\ \bibinfo {pages}
  {2016} (\bibinfo {year} {2017})}\BibitemShut {NoStop}%
\bibitem [{\citenamefont {Heinrich}\ \emph {et~al.}(2015)\citenamefont
  {Heinrich}, \citenamefont {Braun}, \citenamefont {Pascual},\ and\
  \citenamefont {Franke}}]{Heinrich2015}%
  \BibitemOpen
  \bibfield  {author} {\bibinfo {author} {\bibfnamefont {B.~W.}\ \bibnamefont
  {Heinrich}}, \bibinfo {author} {\bibfnamefont {L.}~\bibnamefont {Braun}},
  \bibinfo {author} {\bibfnamefont {J.~I.}\ \bibnamefont {Pascual}}, \ and\
  \bibinfo {author} {\bibfnamefont {K.~J.}\ \bibnamefont {Franke}},\ }\href
  {\doibase 10.1021/acs.nanolett.5b00987} {\bibfield  {journal} {\bibinfo
  {journal} {Nano Lett.}\ }\textbf {\bibinfo {volume} {15}},\ \bibinfo {pages}
  {4024} (\bibinfo {year} {2015})}\BibitemShut {NoStop}%
\bibitem [{\citenamefont {Hiraoka}\ \emph {et~al.}(2017)\citenamefont
  {Hiraoka}, \citenamefont {Minamitami}, \citenamefont {Arafune}, \citenamefont
  {Tsukuhara}, \citenamefont {Watanabe}, \citenamefont {Kawai},\ and\
  \citenamefont {Takagi}}]{Hiraoka2017}%
  \BibitemOpen
  \bibfield  {author} {\bibinfo {author} {\bibfnamefont {R.}~\bibnamefont
  {Hiraoka}}, \bibinfo {author} {\bibfnamefont {E.}~\bibnamefont {Minamitami}},
  \bibinfo {author} {\bibfnamefont {R.}~\bibnamefont {Arafune}}, \bibinfo
  {author} {\bibfnamefont {N.}~\bibnamefont {Tsukuhara}}, \bibinfo {author}
  {\bibfnamefont {S.}~\bibnamefont {Watanabe}}, \bibinfo {author}
  {\bibfnamefont {M.}~\bibnamefont {Kawai}}, \ and\ \bibinfo {author}
  {\bibfnamefont {N.}~\bibnamefont {Takagi}},\ }\href {\doibase
  10.1038/ncomms16012} {\bibfield  {journal} {\bibinfo  {journal} {Nat.
  Commun.}\ }\textbf {\bibinfo {volume} {8}},\ \bibinfo {pages} {16012}
  (\bibinfo {year} {2017})}\BibitemShut {NoStop}%
\bibitem [{\citenamefont {Brand}\ \emph {et~al.}(2018)\citenamefont {Brand},
  \citenamefont {Gozdzik}, \citenamefont {N\'eel}, \citenamefont {Lado},
  \citenamefont {Fern\'andez-Rossier},\ and\ \citenamefont
  {Kr\"oger}}]{Brand2018}%
  \BibitemOpen
  \bibfield  {author} {\bibinfo {author} {\bibfnamefont {J.}~\bibnamefont
  {Brand}}, \bibinfo {author} {\bibfnamefont {S.}~\bibnamefont {Gozdzik}},
  \bibinfo {author} {\bibfnamefont {N.}~\bibnamefont {N\'eel}}, \bibinfo
  {author} {\bibfnamefont {J.~L.}\ \bibnamefont {Lado}}, \bibinfo {author}
  {\bibfnamefont {J.}~\bibnamefont {Fern\'andez-Rossier}}, \ and\ \bibinfo
  {author} {\bibfnamefont {J.}~\bibnamefont {Kr\"oger}},\ }\href {\doibase
  10.1103/PhysRevB.97.195429} {\bibfield  {journal} {\bibinfo  {journal} {Phys.
  Rev. B}\ }\textbf {\bibinfo {volume} {97}},\ \bibinfo {pages} {195429}
  (\bibinfo {year} {2018})}\BibitemShut {NoStop}%
\bibitem [{SM()}]{SM}%
  \BibitemOpen
  \href@noop {} {}\bibinfo {note} {See supplementary material (which includes
  Ref.\cite{Heinrich2013, Heinrich2013b, Pientka2013}) for considerations of
  the relevant parameters, a theoretical description of the change in local
  order parameter and YSR states, including an analytical estimation,
  experimental evidence for the absence of a Cl ligand on FeP, and additional
  experimental data FeP molecules in different environment.}\BibitemShut
  {Stop}%
\bibitem [{\citenamefont {Ruby}\ \emph
  {et~al.}(2015{\natexlab{c}})\citenamefont {Ruby}, \citenamefont {Heinrich},
  \citenamefont {Pascual},\ and\ \citenamefont {Franke}}]{RubyPb15}%
  \BibitemOpen
  \bibfield  {author} {\bibinfo {author} {\bibfnamefont {M.}~\bibnamefont
  {Ruby}}, \bibinfo {author} {\bibfnamefont {B.~W.}\ \bibnamefont {Heinrich}},
  \bibinfo {author} {\bibfnamefont {J.~I.}\ \bibnamefont {Pascual}}, \ and\
  \bibinfo {author} {\bibfnamefont {K.~J.}\ \bibnamefont {Franke}},\ }\href
  {\doibase 10.1103/PhysRevLett.114.157001} {\bibfield  {journal} {\bibinfo
  {journal} {Phys. Rev. Lett.}\ }\textbf {\bibinfo {volume} {114}},\ \bibinfo
  {pages} {157001} (\bibinfo {year} {2015}{\natexlab{c}})}\BibitemShut
  {NoStop}%
\bibitem [{\citenamefont {Bauer}\ \emph {et~al.}(2013)\citenamefont {Bauer},
  \citenamefont {Pascual},\ and\ \citenamefont {Franke}}]{Bauer2013}%
  \BibitemOpen
  \bibfield  {author} {\bibinfo {author} {\bibfnamefont {J.}~\bibnamefont
  {Bauer}}, \bibinfo {author} {\bibfnamefont {J.~I.}\ \bibnamefont {Pascual}},
  \ and\ \bibinfo {author} {\bibfnamefont {K.~J.}\ \bibnamefont {Franke}},\
  }\href {\doibase 10.1103/PhysRevB.87.075125} {\bibfield  {journal} {\bibinfo
  {journal} {Phys. Rev. B}\ }\textbf {\bibinfo {volume} {87}},\ \bibinfo
  {pages} {075125} (\bibinfo {year} {2013})}\BibitemShut {NoStop}%
\bibitem [{\citenamefont {Ali}\ \emph {et~al.}(2012)\citenamefont {Ali},
  \citenamefont {Sanyal},\ and\ \citenamefont {Oppeneer}}]{Ali2012}%
  \BibitemOpen
  \bibfield  {author} {\bibinfo {author} {\bibfnamefont {M.~E.}\ \bibnamefont
  {Ali}}, \bibinfo {author} {\bibfnamefont {B.}~\bibnamefont {Sanyal}}, \ and\
  \bibinfo {author} {\bibfnamefont {P.~M.}\ \bibnamefont {Oppeneer}},\ }\href
  {\doibase 10.1021/jp3021563} {\bibfield  {journal} {\bibinfo  {journal} {J.
  Phys. Chem. B}\ }\textbf {\bibinfo {volume} {116}},\ \bibinfo {pages} {5849}
  (\bibinfo {year} {2012})}\BibitemShut {NoStop}%
\bibitem [{\citenamefont {Bhandary}\ \emph {et~al.}(2016)\citenamefont
  {Bhandary}, \citenamefont {Sch{\"{u}}ler}, \citenamefont {Thunstr{\"{o}}m},
  \citenamefont {{Di Marco}}, \citenamefont {Brena}, \citenamefont {Eriksson},
  \citenamefont {Wehling},\ and\ \citenamefont {Sanyal}}]{Bhandary2016}%
  \BibitemOpen
  \bibfield  {author} {\bibinfo {author} {\bibfnamefont {S.}~\bibnamefont
  {Bhandary}}, \bibinfo {author} {\bibfnamefont {M.}~\bibnamefont
  {Sch{\"{u}}ler}}, \bibinfo {author} {\bibfnamefont {P.}~\bibnamefont
  {Thunstr{\"{o}}m}}, \bibinfo {author} {\bibfnamefont {I.}~\bibnamefont {{Di
  Marco}}}, \bibinfo {author} {\bibfnamefont {B.}~\bibnamefont {Brena}},
  \bibinfo {author} {\bibfnamefont {O.}~\bibnamefont {Eriksson}}, \bibinfo
  {author} {\bibfnamefont {T.}~\bibnamefont {Wehling}}, \ and\ \bibinfo
  {author} {\bibfnamefont {B.}~\bibnamefont {Sanyal}},\ }\href {\doibase
  10.1103/PhysRevB.93.155158} {\bibfield  {journal} {\bibinfo  {journal} {Phys.
  Rev. B}\ }\textbf {\bibinfo {volume} {93}},\ \bibinfo {pages} {155158}
  (\bibinfo {year} {2016})}\BibitemShut {NoStop}%
\bibitem [{\citenamefont {Rubio-Verd{\'{u}}}\ \emph {et~al.}(2018)\citenamefont
  {Rubio-Verd{\'{u}}}, \citenamefont {Sarasola}, \citenamefont {Choi},
  \citenamefont {Majzik}, \citenamefont {Ebeling}, \citenamefont {Calvo},
  \citenamefont {Ugeda}, \citenamefont {Garcia-Lekue}, \citenamefont
  {S{\'{a}}nchez-Portal},\ and\ \citenamefont {Pascual}}]{Rubio-Verdu2018}%
  \BibitemOpen
  \bibfield  {author} {\bibinfo {author} {\bibfnamefont {C.}~\bibnamefont
  {Rubio-Verd{\'{u}}}}, \bibinfo {author} {\bibfnamefont {A.}~\bibnamefont
  {Sarasola}}, \bibinfo {author} {\bibfnamefont {D.-J.}\ \bibnamefont {Choi}},
  \bibinfo {author} {\bibfnamefont {Z.}~\bibnamefont {Majzik}}, \bibinfo
  {author} {\bibfnamefont {R.}~\bibnamefont {Ebeling}}, \bibinfo {author}
  {\bibfnamefont {M.~R.}\ \bibnamefont {Calvo}}, \bibinfo {author}
  {\bibfnamefont {M.~M.}\ \bibnamefont {Ugeda}}, \bibinfo {author}
  {\bibfnamefont {A.}~\bibnamefont {Garcia-Lekue}}, \bibinfo {author}
  {\bibfnamefont {D.}~\bibnamefont {S{\'{a}}nchez-Portal}}, \ and\ \bibinfo
  {author} {\bibfnamefont {J.~I.}\ \bibnamefont {Pascual}},\ }\href {\doibase
  10.1038/s42005-018-0015-6} {\bibfield  {journal} {\bibinfo  {journal}
  {Commun. Phys.}\ }\textbf {\bibinfo {volume} {1}},\ \bibinfo {pages} {15}
  (\bibinfo {year} {2018})}\BibitemShut {NoStop}%
\bibitem [{\citenamefont {Randeria}\ \emph {et~al.}(2016)\citenamefont
  {Randeria}, \citenamefont {Feldman}, \citenamefont {Drozdov},\ and\
  \citenamefont {Yazdani}}]{Randeria2016}%
  \BibitemOpen
  \bibfield  {author} {\bibinfo {author} {\bibfnamefont {M.~T.}\ \bibnamefont
  {Randeria}}, \bibinfo {author} {\bibfnamefont {B.~E.}\ \bibnamefont
  {Feldman}}, \bibinfo {author} {\bibfnamefont {I.~K.}\ \bibnamefont
  {Drozdov}}, \ and\ \bibinfo {author} {\bibfnamefont {A.}~\bibnamefont
  {Yazdani}},\ }\href {\doibase 10.1103/PhysRevB.93.161115} {\bibfield
  {journal} {\bibinfo  {journal} {Phys. Rev. B}\ }\textbf {\bibinfo {volume}
  {93}},\ \bibinfo {pages} {161115} (\bibinfo {year} {2016})}\BibitemShut
  {NoStop}%
\bibitem [{\citenamefont {Ternes}\ \emph {et~al.}(2006)\citenamefont {Ternes},
  \citenamefont {Schneider}, \citenamefont {Cuevas}, \citenamefont {Lutz},
  \citenamefont {Hirjibehedin},\ and\ \citenamefont {Heinrich}}]{Ternes2006}%
  \BibitemOpen
  \bibfield  {author} {\bibinfo {author} {\bibfnamefont {M.}~\bibnamefont
  {Ternes}}, \bibinfo {author} {\bibfnamefont {W.~D.}\ \bibnamefont
  {Schneider}}, \bibinfo {author} {\bibfnamefont {J.~C.}\ \bibnamefont
  {Cuevas}}, \bibinfo {author} {\bibfnamefont {C.~P.}\ \bibnamefont {Lutz}},
  \bibinfo {author} {\bibfnamefont {C.~F.}\ \bibnamefont {Hirjibehedin}}, \
  and\ \bibinfo {author} {\bibfnamefont {A.~J.}\ \bibnamefont {Heinrich}},\
  }\href {\doibase 10.1103/PhysRevB.74.132501} {\bibfield  {journal} {\bibinfo
  {journal} {Phys. Rev. B}\ }\textbf {\bibinfo {volume} {74}},\ \bibinfo
  {pages} {132501} (\bibinfo {year} {2006})}\BibitemShut {NoStop}%
\bibitem [{\citenamefont {Andersen}\ \emph {et~al.}(2011)\citenamefont
  {Andersen}, \citenamefont {Flensberg}, \citenamefont {Koerting},\ and\
  \citenamefont {Paaske}}]{Andersen2011}%
  \BibitemOpen
  \bibfield  {author} {\bibinfo {author} {\bibfnamefont {B.~M.}\ \bibnamefont
  {Andersen}}, \bibinfo {author} {\bibfnamefont {K.}~\bibnamefont {Flensberg}},
  \bibinfo {author} {\bibfnamefont {V.}~\bibnamefont {Koerting}}, \ and\
  \bibinfo {author} {\bibfnamefont {J.}~\bibnamefont {Paaske}},\ }\href
  {\doibase 10.1103/PhysRevLett.107.256802} {\bibfield  {journal} {\bibinfo
  {journal} {Phys. Rev. Lett.}\ }\textbf {\bibinfo {volume} {107}},\ \bibinfo
  {pages} {256802} (\bibinfo {year} {2011})}\BibitemShut {NoStop}%
\bibitem [{\citenamefont {Heinrich}\ \emph
  {et~al.}(2013{\natexlab{a}})\citenamefont {Heinrich}, \citenamefont {Ahmadi},
  \citenamefont {Müller}, \citenamefont {Braun}, \citenamefont {Pascual},\
  and\ \citenamefont {Franke}}]{Heinrich2013}%
  \BibitemOpen
  \bibfield  {author} {\bibinfo {author} {\bibfnamefont {B.~W.}\ \bibnamefont
  {Heinrich}}, \bibinfo {author} {\bibfnamefont {G.}~\bibnamefont {Ahmadi}},
  \bibinfo {author} {\bibfnamefont {V.~L.}\ \bibnamefont {Müller}}, \bibinfo
  {author} {\bibfnamefont {L.}~\bibnamefont {Braun}}, \bibinfo {author}
  {\bibfnamefont {J.~I.}\ \bibnamefont {Pascual}}, \ and\ \bibinfo {author}
  {\bibfnamefont {K.~J.}\ \bibnamefont {Franke}},\ }\href {\doibase
  10.1021/nl402575c} {\bibfield  {journal} {\bibinfo  {journal} {Nano Lett.}\
  }\textbf {\bibinfo {volume} {13}},\ \bibinfo {pages} {4840} (\bibinfo {year}
  {2013}{\natexlab{a}})}\BibitemShut {NoStop}%
\bibitem [{\citenamefont {Heinrich}\ \emph
  {et~al.}(2013{\natexlab{b}})\citenamefont {Heinrich}, \citenamefont {Braun},
  \citenamefont {Pascual},\ and\ \citenamefont {Franke}}]{Heinrich2013b}%
  \BibitemOpen
  \bibfield  {author} {\bibinfo {author} {\bibfnamefont {B.~W.}\ \bibnamefont
  {Heinrich}}, \bibinfo {author} {\bibfnamefont {L.}~\bibnamefont {Braun}},
  \bibinfo {author} {\bibfnamefont {J.~I.}\ \bibnamefont {Pascual}}, \ and\
  \bibinfo {author} {\bibfnamefont {K.~J.}\ \bibnamefont {Franke}},\ }\href
  {\doibase 10.1038/nphys2794} {\bibfield  {journal} {\bibinfo  {journal} {Nat.
  Phys.}\ }\textbf {\bibinfo {volume} {9}},\ \bibinfo {pages} {765} (\bibinfo
  {year} {2013}{\natexlab{b}})}\BibitemShut {NoStop}%
\bibitem [{\citenamefont {Pientka}\ \emph {et~al.}(2013)\citenamefont
  {Pientka}, \citenamefont {Glazman},\ and\ \citenamefont {von
  Oppen}}]{Pientka2013}%
  \BibitemOpen
  \bibfield  {author} {\bibinfo {author} {\bibfnamefont {F.}~\bibnamefont
  {Pientka}}, \bibinfo {author} {\bibfnamefont {L.~I.}\ \bibnamefont
  {Glazman}}, \ and\ \bibinfo {author} {\bibfnamefont {F.}~\bibnamefont {von
  Oppen}},\ }\href {\doibase 10.1103/PhysRevB.88.155420} {\bibfield  {journal}
  {\bibinfo  {journal} {Phys. Rev. B}\ }\textbf {\bibinfo {volume} {88}},\
  \bibinfo {pages} {155420} (\bibinfo {year} {2013})}\BibitemShut {NoStop}%
\end{thebibliography}

\begin{thebibliography}{10}
\makeatletter
\providecommand \@ifxundefined [1]{%
 \@ifx{#1\undefined}
}%
\providecommand \@ifnum [1]{%
 \ifnum #1\expandafter \@firstoftwo
 \else \expandafter \@secondoftwo
 \fi
}%
\providecommand \@ifx [1]{%
 \ifx #1\expandafter \@firstoftwo
 \else \expandafter \@secondoftwo
 \fi
}%
\providecommand \natexlab [1]{#1}%
\providecommand \enquote  [1]{``#1''}%
\providecommand \bibnamefont  [1]{#1}%
\providecommand \bibfnamefont [1]{#1}%
\providecommand \citenamefont [1]{#1}%
\providecommand \href@noop [0]{\@secondoftwo}%
\providecommand \href [0]{\begingroup \@sanitize@url \@href}%
\providecommand \@href[1]{\@@startlink{#1}\@@href}%
\providecommand \@@href[1]{\endgroup#1\@@endlink}%
\providecommand \@sanitize@url [0]{\catcode `\\12\catcode `\$12\catcode
  `\&12\catcode `\#12\catcode `\^12\catcode `\_12\catcode `\%12\relax}%
\providecommand \@@startlink[1]{}%
\providecommand \@@endlink[0]{}%
\providecommand \url  [0]{\begingroup\@sanitize@url \@url }%
\providecommand \@url [1]{\endgroup\@href {#1}{\urlprefix }}%
\providecommand \urlprefix  [0]{URL }%
\providecommand \Eprint [0]{\href }%
\providecommand \doibase [0]{http://dx.doi.org/}%
\providecommand \selectlanguage [0]{\@gobble}%
\providecommand \bibinfo  [0]{\@secondoftwo}%
\providecommand \bibfield  [0]{\@secondoftwo}%
\providecommand \translation [1]{[#1]}%
\providecommand \BibitemOpen [0]{}%
\providecommand \bibitemStop [0]{}%
\providecommand \bibitemNoStop [0]{.\EOS\space}%
\providecommand \EOS [0]{\spacefactor3000\relax}%
\providecommand \BibitemShut  [1]{\csname bibitem#1\endcsname}%
\let\auto@bib@innerbib\@empty
\bibitem [{\citenamefont {Balatsky}\ \emph {et~al.}(2006)\citenamefont
  {Balatsky}, \citenamefont {Vekhter},\ and\ \citenamefont
  {Zhu}}]{SBalatsky2006}%
  \BibitemOpen
  \bibfield  {author} {\bibinfo {author} {\bibfnamefont {A.~V.}\ \bibnamefont
  {Balatsky}}, \bibinfo {author} {\bibfnamefont {I.}~\bibnamefont {Vekhter}}, \
  and\ \bibinfo {author} {\bibfnamefont {J.~X.}\ \bibnamefont {Zhu}},\ }\href
  {\doibase 10.1103/RevModPhys.78.373} {\bibfield  {journal} {\bibinfo
  {journal} {Rev. Mod. Phys.}\ }\textbf {\bibinfo {volume} {78}},\ \bibinfo
  {pages} {373} (\bibinfo {year} {2006})}\BibitemShut {NoStop}%
\bibitem [{\citenamefont {Ruby}\ \emph {et~al.}(2015)\citenamefont {Ruby},
  \citenamefont {Pientka}, \citenamefont {Peng}, \citenamefont {von Oppen},
  \citenamefont {Heinrich},\ and\ \citenamefont {Franke}}]{SRuby2015}%
  \BibitemOpen
  \bibfield  {author} {\bibinfo {author} {\bibfnamefont {M.}~\bibnamefont
  {Ruby}}, \bibinfo {author} {\bibfnamefont {F.}~\bibnamefont {Pientka}},
  \bibinfo {author} {\bibfnamefont {Y.}~\bibnamefont {Peng}}, \bibinfo {author}
  {\bibfnamefont {F.}~\bibnamefont {von Oppen}}, \bibinfo {author}
  {\bibfnamefont {B.~W.}\ \bibnamefont {Heinrich}}, \ and\ \bibinfo {author}
  {\bibfnamefont {K.~J.}\ \bibnamefont {Franke}},\ }\href {\doibase
  10.1103/PhysRevLett.115.087001} {\bibfield  {journal} {\bibinfo  {journal}
  {Phys. Rev. Lett.}\ }\textbf {\bibinfo {volume} {115}},\ \bibinfo {pages}
  {087001} (\bibinfo {year} {2015})}\BibitemShut {NoStop}%
\bibitem [{\citenamefont {Pientka}\ \emph {et~al.}(2013)\citenamefont
  {Pientka}, \citenamefont {Glazman},\ and\ \citenamefont {von
  Oppen}}]{SPientka2013}%
  \BibitemOpen
  \bibfield  {author} {\bibinfo {author} {\bibfnamefont {F.}~\bibnamefont
  {Pientka}}, \bibinfo {author} {\bibfnamefont {L.~I.}\ \bibnamefont
  {Glazman}}, \ and\ \bibinfo {author} {\bibfnamefont {F.}~\bibnamefont {von
  Oppen}},\ }\href {\doibase 10.1103/PhysRevB.88.155420} {\bibfield  {journal}
  {\bibinfo  {journal} {Phys. Rev. B}\ }\textbf {\bibinfo {volume} {88}},\
  \bibinfo {pages} {155420} (\bibinfo {year} {2013})}\BibitemShut {NoStop}%
\bibitem [{\citenamefont {Salkola}\ \emph {et~al.}(1997)\citenamefont
  {Salkola}, \citenamefont {Balatsky},\ and\ \citenamefont
  {Schrieffer}}]{SSalkola1997}%
  \BibitemOpen
  \bibfield  {author} {\bibinfo {author} {\bibfnamefont {M.~I.}\ \bibnamefont
  {Salkola}}, \bibinfo {author} {\bibfnamefont {a.~V.}\ \bibnamefont
  {Balatsky}}, \ and\ \bibinfo {author} {\bibfnamefont {J.~R.}\ \bibnamefont
  {Schrieffer}},\ }\href {\doibase 10.1103/PhysRevB.55.12648} {\bibfield
  {journal} {\bibinfo  {journal} {Phys. Rev. B}\ }\textbf {\bibinfo {volume}
  {55}},\ \bibinfo {pages} {12648} (\bibinfo {year} {1997})}\BibitemShut
  {NoStop}%
\bibitem [{\citenamefont {Flatt{\'{e}}}\ and\ \citenamefont
  {Byers}(1997)}]{SFlatte1997}%
  \BibitemOpen
  \bibfield  {author} {\bibinfo {author} {\bibfnamefont {M.~E.}\ \bibnamefont
  {Flatt{\'{e}}}}\ and\ \bibinfo {author} {\bibfnamefont {J.~M.}\ \bibnamefont
  {Byers}},\ }\href {\doibase https://doi.org/10.1103/PhysRevB.56.11213}
  {\bibfield  {journal} {\bibinfo  {journal} {Phys. Rev. B}\ }\textbf {\bibinfo
  {volume} {56}},\ \bibinfo {pages} {11213} (\bibinfo {year}
  {1997})}\BibitemShut {NoStop}%
\bibitem [{\citenamefont {Meng}\ \emph {et~al.}(2015)\citenamefont {Meng},
  \citenamefont {Klinovaja}, \citenamefont {Hoffman}, \citenamefont {Simon},\
  and\ \citenamefont {Loss}}]{SMeng2015}%
  \BibitemOpen
  \bibfield  {author} {\bibinfo {author} {\bibfnamefont {T.}~\bibnamefont
  {Meng}}, \bibinfo {author} {\bibfnamefont {J.}~\bibnamefont {Klinovaja}},
  \bibinfo {author} {\bibfnamefont {S.}~\bibnamefont {Hoffman}}, \bibinfo
  {author} {\bibfnamefont {P.}~\bibnamefont {Simon}}, \ and\ \bibinfo {author}
  {\bibfnamefont {D.}~\bibnamefont {Loss}},\ }\href {\doibase
  10.1103/PhysRevB.92.064503} {\bibfield  {journal} {\bibinfo  {journal} {Phys.
  Rev. B}\ }\textbf {\bibinfo {volume} {92}},\ \bibinfo {pages} {064503}
  (\bibinfo {year} {2015})}\BibitemShut {NoStop}%
\bibitem [{\citenamefont {Heinrich}\ \emph
  {et~al.}(2013{\natexlab{a}})\citenamefont {Heinrich}, \citenamefont {Ahmadi},
  \citenamefont {Müller}, \citenamefont {Braun}, \citenamefont {Pascual},\
  and\ \citenamefont {Franke}}]{SHeinrich2013}%
  \BibitemOpen
  \bibfield  {author} {\bibinfo {author} {\bibfnamefont {B.~W.}\ \bibnamefont
  {Heinrich}}, \bibinfo {author} {\bibfnamefont {G.}~\bibnamefont {Ahmadi}},
  \bibinfo {author} {\bibfnamefont {V.~L.}\ \bibnamefont {Müller}}, \bibinfo
  {author} {\bibfnamefont {L.}~\bibnamefont {Braun}}, \bibinfo {author}
  {\bibfnamefont {J.~I.}\ \bibnamefont {Pascual}}, \ and\ \bibinfo {author}
  {\bibfnamefont {K.~J.}\ \bibnamefont {Franke}},\ }\href {\doibase
  10.1021/nl402575c} {\bibfield  {journal} {\bibinfo  {journal} {Nano Lett.}\
  }\textbf {\bibinfo {volume} {13}},\ \bibinfo {pages} {4840} (\bibinfo {year}
  {2013}{\natexlab{a}})}\BibitemShut {NoStop}%
\bibitem [{\citenamefont {Heinrich}\ \emph
  {et~al.}(2013{\natexlab{b}})\citenamefont {Heinrich}, \citenamefont {Braun},
  \citenamefont {Pascual},\ and\ \citenamefont {Franke}}]{SHeinrich2013b}%
  \BibitemOpen
  \bibfield  {author} {\bibinfo {author} {\bibfnamefont {B.~W.}\ \bibnamefont
  {Heinrich}}, \bibinfo {author} {\bibfnamefont {L.}~\bibnamefont {Braun}},
  \bibinfo {author} {\bibfnamefont {J.~I.}\ \bibnamefont {Pascual}}, \ and\
  \bibinfo {author} {\bibfnamefont {K.~J.}\ \bibnamefont {Franke}},\ }\href
  {\doibase 10.1038/nphys2794} {\bibfield  {journal} {\bibinfo  {journal} {Nat.
  Phys.}\ }\textbf {\bibinfo {volume} {9}},\ \bibinfo {pages} {765} (\bibinfo
  {year} {2013}{\natexlab{b}})}\BibitemShut {NoStop}%
	\end{thebibliography}

%

\clearpage

\setcounter{figure}{0}
\setcounter{section}{0}
\setcounter{equation}{0}
\renewcommand{\theequation}{S\arabic{equation}}
\renewcommand{\thefigure}{S\arabic{figure}}

\onecolumngrid

\renewcommand{\Fig}[1]{\mbox{Fig.\unitspace\ref{Sfig:#1}}}
\renewcommand{\Figure}[1]{\mbox{Figure\unitspace\ref{Sfig:#1}}}

\newcommand{\vsigma}{\mbox{\boldmath $\sigma$}}

\section*{\Large{Supplemental Material}}

\vspace{0.7cm}
\section{Theoretical considerations}

In this section, we summarize relevant theoretical considerations to understand the experimental results, including considerations of relevant parameters for driving the quantum phase transition, new analytical results on the effect of self-consistency on the local gap parameter $\Delta({\bf r})$, and the energy $\epsilon$ of the YSR state. 

\subsection{Relation between experimental and model parameters}

A minimal model for the scattering of substrate electrons by a magnetic impurity is the $s-d$ model (${\cal H} = (V + J {\bf S}\cdot{\vsigma})$) which includes the hybridization between the impurity orbitals and the substrate through the exchange coupling  $J$ and the potential scattering $V$ (as obtained by a Schrieffer-Wolff transformation from the Anderson model). Treating the impurity spin ${\bf S}$ as classical, the energies of the YSR states are given by
\begin{equation} \label{YSR}
\epsilon=\pm \Delta \left(\frac{1-\alpha^2+\beta^2}{\sqrt{\left(1-\alpha^2+\beta^2\right)^2+4\alpha^2}}\right)
\end{equation}
with $\alpha=\pi \rho_0 SJ$ and $\beta=\pi \rho_0 V$. Here, $\rho_0$ is the normal-state density of states of the superconductor. As shown in the main manuscript, the approach of the STM tip to the FeP molecule shifts the YSR states substantially within the superconducting gap. Specifically, it shifts the YSR energies  across the Fermi level, corresponding to the quantum phase transition. While according to Eq.\ \ref{YSR}, the YSR energies depend on  several parameters, we argue that the exchange coupling strength $J$ is the most relevant parameter for driving the quantum phase transition. In the following we explain why the other parameters are expected to affect the YSR energies only slightly, and cannot drive the quantum phase transition by themselves.

{\em	 Exchange coupling} $J$: The tip approach towards the molecule locally modifies the molecule-substrate distance as a result of tip-molecule forces. This changes the molecule-substrate hybridization and hence the exchange coupling $J$. As can be seen from Eq.\ (\ref{YSR}), $J$ is essential for changing the sign of the YSR energy and, hence, the quantum phase transition. Moreover, all qualitiative features of our experimental data can be captured by the variation of this parameter in response to the force field of the STM tip, which induces conformational changes of the molecule.

{\em Potential scattering} $V$: The molecule-substrate hybridization also induces the potential scattering amplitude $V$. As $V$ (and hence $\beta$) depends on the relative position of the singly and doubly occupied orbitals with respect to the Fermi level, $V$ also accounts for changes in the charge transfer between impurity and substrate. However unlike $J$, the potential scattering $V$ does not grow under renormalization group transformations of the $s-d$ model, so that $V$ is generically expected to be weaker than $J$.

{\em Normal density of states} $\rho_0$ {\em and gap parameter} $\Delta$: One can safely assume that $\rho_0$ cannot drive the quantum phase transition as it is a bulk parameter. Self-consistent calculations predict a $J$-dependent change in $\Delta$ at the impurity site. However, we show explicitly in the next section that this modification of $\Delta$ causes only  minimal shifts of the YSR state energy which are substantially below the energy resolution of the experiment.

{\em 	Impurity spin} $S$:
The tip approach may also modify the magnetic moment and thus the effective spin. However, we expect that the spin state is only changed by a small fraction of the total spin moment. If the spin magnetic moment changed by more than $\Delta S=1/2$, we would expect more substantial changes in the subgap spectrum such as a variation in the number of YSR states. A small change is insufficient to explain the substantial shift in the YSR state energy.

While some of these additional parameters may thus be relevant for explaining quantitative details of the experiment, it suffices to focus on the exchange coupling $J$ to understand the qualitiative features of the experimental data.  This is the approach which we take in the main manuscript.

\subsection{Gap parameter and YSR state energy near the quantum phase transition}
\label{secYSR}

In the absence of the magnetic impurity, the superconducting ground state involves an even number of electrons (even fermion number parity). Excited states with odd numbers of electrons (odd fermion parity) have a continuous spectrum with an excitation energy larger than the superconducting gap $\Delta$. A magnetic impurity induces a localized quasiparticle state at subgap energies \cite{SBalatsky2006}. 

For weak exchange coupling between impurity and electrons in the superconductor, this YSR state is unoccupied in the ground state so that the ground state remains fully paired (even fermion parity). But there is now a discrete excited state with odd fermion parity at an excitation energy $\epsilon$ below $\Delta$ in which the YSR state is occupied by a quasiparticle. 

The excitation energy of this odd fermion parity state reduces with increasing exchange coupling until it reaches zero at a critical coupling $J_c$. Beyond the critical coupling, the bound quasiparticle state becomes occupied in the ground state and empty in the excited state, i.e., the ground state is now an odd fermion parity state while the excited state has even parity. This level crossing between states with even and odd fermion parity at the critical coupling is a true level crossing as it is protected by fermion parity conservation
\cite{SBalatsky2006}. 

At zero temperature, tunneling into superconductors at subgap energies proceeds via Andreev processes which transfer Cooper pairs into the superconductor. At finite temperatures, inelastic processes open an additional tunneling channel in which single electrons are transferred. An electron tunnels into the superconductor occupying a subgap quasiparticle state and is subsequently excited inelastically into the quasiparticle continuum (or recombines inelastically into a Cooper pair with another subgap excitation). This single-particle tunneling dominates over Andreev processes when the tunneling rate is slow compared to the inelastic relaxation rate. This is the case in STM experiments at sufficiently low tunneling conductance
\cite{SRuby2015}. 

We start from a $4\times4$ Bogoliubov-deGennes (BdG) Hamiltonian 
\begin{equation}
  {\cal H} = \xi_{\bf p}\tau_z  +(V\tau_z + J {\bf S}_i\cdot{\vsigma}) \delta({\bf r})  + \Delta \tau_x.
  \label{Ham}
\end{equation}
in the basis in which the four-component Nambu operator takes the form $\Psi = [\psi_\uparrow, \psi_\downarrow, \psi^\dagger_\downarrow, -\psi^\dagger_\uparrow]$ in terms of the electronic field operator $\psi_\sigma({\bf r})$ and, for definiteness, consider a classical magnetic impurity. Here, $\xi_{\bf p} = {\bf p}^2/2m-\mu$ with the chemical potential $\mu$, $V$ the strength of the potential scattering by the impurity, and $J$ denotes the strength of the exchange coupling between the magnetic impurity with spin $S$ and the electrons in the superconductor. The Pauli matrices $\sigma_i$ ($\tau_i$) operate in spin (particle-hole) space. Choosing the impurity spin ${\bf S}$ to point along the $z$ direction, the $4\times 4$ Hamiltonian separates into independent $2\times2$ Hamiltonians 
\begin{equation}
 {\cal H}_\pm = \xi_{\bf p} \tau_z (V\tau_z \pm J S) \delta({\bf r}) + \Delta \tau_x.
 \label{Hpm}
\end{equation}
A standard calculation \cite{SPientka2013} shows that $H_+$ ($H_-$) has one subgap solution $[u_\epsilon({\bf r}),v_\epsilon({\bf r}]$ ($[u_{-\epsilon}({\bf r}),v_{-\epsilon}({\bf r})]$), whose energy we denote by $\epsilon$ ($-\epsilon$). Since the BdG formalism doubles the degrees of freedom, it is sufficient to consider only the solutions of one of these Hamiltonians, say $H_-$. Its subgap (YSR) state starts out at positive energies at small exchange couplings and crosses to negative energies for strong coupling. Keeping only the contribution of the subgap state with Bogoliubov operator 
\begin{equation}
  \gamma_\epsilon = \int d{\bf r} [u^*_\epsilon({\bf r})  \psi_\downarrow({\bf r}) - v^*_\epsilon({\bf r})  \psi_\uparrow^\dagger({\bf r}) ]     ,
  \end{equation}
we can write the electron operators as
\begin{equation}
  \psi_\downarrow({\bf r}) = u_\epsilon({\bf r}) \gamma_\epsilon+\ldots
\end{equation} 
and 
\begin{equation}
  \psi_\uparrow({\bf r}) = -v^*_\epsilon({\bf r})\gamma_\epsilon^\dagger + \ldots,
\end{equation} 
where the ellipses indicate the contributions of above-gap quasiparticles. The even parity state satisfies 
$\gamma_\epsilon|{\rm even}\rangle =0$ and the odd-parity state is $|{\rm odd}\rangle = \gamma_\epsilon^\dagger|{\rm even}\rangle$. Note that $\gamma_\epsilon^\dagger$ removes an electron spin of 1/2 from the electron system, so that the spin of the odd parity state is lower by 1/2 compared to the (spinless) even parity state. 

In the regime of single-electron tunneling, the differential conductance at $eV=\pm (\Delta_t+\epsilon)$ (for a superconducting tip with gap $\Delta_t$)  provides access to the Bogoliubov-deGennes wave function of the YSR bound state (energy $\epsilon$) at the tip position ${\bf r}$ \cite{SRuby2015}. Consider first a positive bias voltage with electrons tunneling into the superconducting substrate. 
\begin{itemize}
\item For $J<J_c$, the tunneling electrons excite the system from the even parity ground state to the odd parity excited state. In view of the spin polarization of the YSR state, the differential conductance is proportional to $|\langle {\rm odd}|\psi_\downarrow^\dagger({\bf r})|{\rm even}\rangle|^2$  (Fermi's Golden Rule). We therefore deduce that the differential conductance is proportional to $|u({\bf r})|^2$ in terms of the electron component of the YSR state.
\item For $J>J_c$, the differential conductance involves the matrix element $|\langle {\rm even}|\psi_\uparrow^\dagger({\bf r})|{\rm odd}\rangle|^2$, since $|{\rm odd}\rangle$ ($|{\rm even}\rangle$) is now the ground (excited) state. The differential conductance is thus proportional to $|v({\bf r})|^2$ involving the hole wave function of the bound state. 
\end{itemize}
As the electron and hole wave functions are generically different, one expects a discontinuous change in the differential conductance at the critical coupling $J_c$. 

Now consider negative bias voltages where electrons are tunneling out of the superconducting substrate. The differential conductance is then proportional to   $|\langle {\rm odd}|\psi_\uparrow({\bf r})|{\rm even}\rangle|^2$ for $J<J_c$ and $|\langle {\rm even}|\psi_\downarrow({\bf r})|{\rm odd}\rangle|^2$ for $J>J_c$. Consequently, the roles of $u$ and $v$ invert relative to positive bias voltages.

This picture of the quantum phase transition must be amended when effects of self-consistency are taken into account. Both the local gap parameter and the energy of the YSR state have discontinuities at the critical coupling. This has been investigated repeatedly by numerical simulations \cite{SSalkola1997,Flatte1997,Meng2015}, but to the best of our knowledge, there are no analytical estimates for the magnitudes of these discontinuities in the literature. Such estimates are provided in the following. 
 
The jump in the local gap parameter can be understood when writing the (zero-temperature) gap equation in terms of the Bogoliubov-deGennes wave functions $u_n({\bf r})$ and $v_n({\bf r})$ of, say, $H_-$,
\begin{equation}
   \Delta({\bf r}) = g \sum_n u_n({\bf r})v_n^*({\bf r}),
\end{equation} 
where the sum is over positive-energy eigenstates within an energy band about the Fermi energy whose width is given by the Debye frequency $\omega_D$. The parameter $g>0$ denotes the coupling constant of the attractive interaction. As the system passes through the quantum phase transition, the positive energy subgap state crosses to negative energies and no longer contributes to the sum on the right hand side of the gap equation. Thus, we find
\begin{equation}
  \delta \Delta({\bf r}) = -g u_{\epsilon}({\bf r})v_{\epsilon}^*({\bf r}).
\end{equation} 
for the jump in the gap parameter. Here, we assume that the effect of the discontinuity on the other Bogoliubov-deGennes eigenstates is weak. We will confirm below that this is indeed the case. For a classical magnetic impurity and uniform gap parameter, one finds that $u_{\epsilon}$ and $v_{\epsilon}$ can be chosen real and have the same sign at the impurity position \cite{SPientka2013}. Thus, this indeed describes a local suppression of the order parameter. 

The discontinuity in the gap parameter is limited to the region in which the YSR bound state is localized. At its center, say ${\bf r}=0$, and at the critical exchange coupling $J_c$, the YSR bound state has an amplitude of $|u(0)|^2\sim |v(0)|^2 \sim \nu_0\Delta$ \cite{SPientka2013}, where $\nu_0$ is the (normal-state) density of states of the superconductor at the Fermi energy. Inserting this into the expression for the jump in the gap parameter, we have $\delta\Delta(0)\sim -g\nu_0\Delta$. Using the relation 
\begin{equation}
   \Delta \sim \omega_D e^{-1/g\nu_0},
\end{equation}
we therefore obtain
\begin{equation}
   \delta \Delta(0) \sim  -\frac{\Delta}{\ln(\omega_D/\Delta)}
\end{equation}
for the local and discontinuous reduction in the gap parameter at the critical coupling. 

As the gap is locally suppressed at the quantum phase transition, it can even change sign. This happens when the discontinuity is larger than the value of the local gap parameter just before the quantum phase transition. For  $J<J_c$, the magnetic impurity suppresses the gap parameter continuously and this suppression of the gap parameter has been previously estimated  within perturbation theory \cite{SMeng2015}, yielding a suppression by $\alpha^2\Delta$ for small exchange coupling $\alpha=\pi\nu_0JS$. As $\alpha\sim 1$ at the quantum phase transition, the value of $\Delta({\bf 0})$ just before the quantum phase transition may indeed be smaller than the magnitude $\delta\Delta(0)$ of the jump, so that the gap parameter changes sign across the transition. 

Since the jump in $\Delta({\bf r})$ is associated with the bound state contribution to the gap equation, it is localized in the vicinity of the impurity. To provide an estimate for the associated jump in the energy of the YSR state, we approximate the jump in $\Delta({\bf r})$ by a $\delta$-function \cite{SMeng2015}, 
\begin{equation}
   \delta\Delta({\bf r}) = a^3 \delta\Delta(0)\delta({\bf r}).
\end{equation}
Here, $a$ is the linear dimension of the region over which the gap parameter is suppressed, which can be estimated as \cite{SFlatte1997} $a\approx 2/k_F$ in terms of the Fermi wave vector of the superconductor. The associated jump in the energy of the YSR state can then be obtained from first-order perturbation theory which yields
\begin{equation}
\delta \epsilon \sim \frac{\Delta^2}{E_F \ln(\omega_D/\Delta)}.
\end{equation}
In deriving this expression, we have again used the magnitude of the YSR wave function at the position of the impurity. This expression shows that the effect of the gap reduction on the Shiba states is small in the parameter $\Delta/E_F$ which is of order $10^{-5}$ for conventional superconductors. 

Clearly, this predicts a discontinuity in the energy of the Shiba state which is below experimental resolution by orders of magnitude. Moreover, in the regime of single electron tunneling, STM experiments are sensitive to the change in the local gap parameter only via the associated change in the bound state wave function which is negligible. 

\section{Additional experimental data}

\subsection{Fe(III)-porphine-chloride molecules}

\begin{figure}[htp]
	\includegraphics [width=0.95\textwidth,clip=]{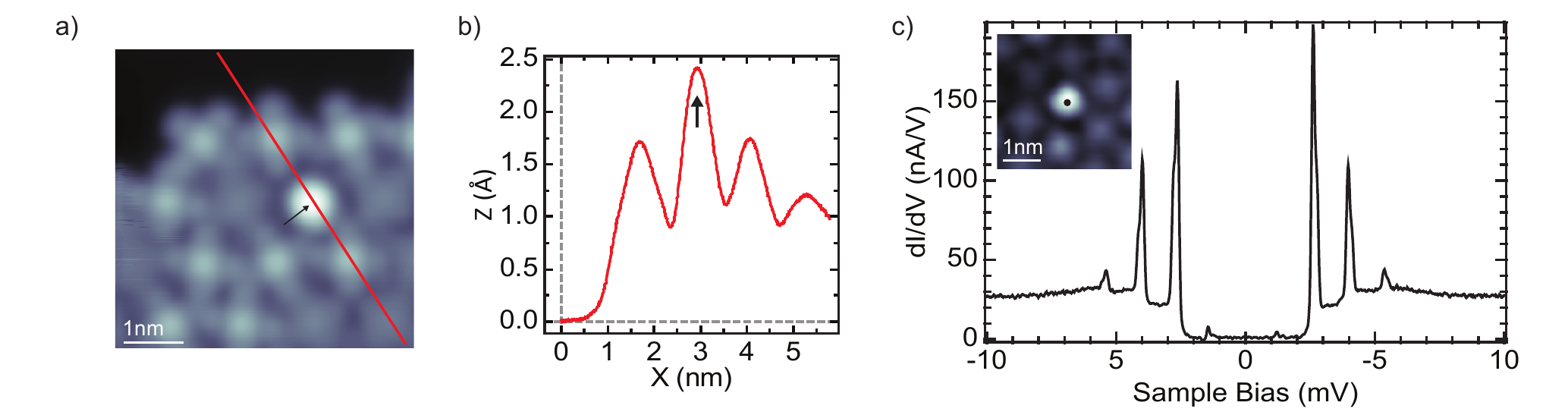}
	\caption{a) STM topography image ($V=-45\umV$, $I=100\upA$) after deposition of FeP-Cl on a sample held below room temperature. A few molecules kept their Cl ligand and show a protrusion above their center. b) Line profile along the red line shown in a) showing the larger apparent height of the molecule indicated by an arrow. c) \didv spectrum acquired above the center of such a molecule (feedback opened at $V=-45\umV$, $I=100\upA$, $V_{\mathrm{rms}}=50\umuV$) showing the presence of inelastic spin excitations in accordance with \cite{SHeinrich2013b}.}
	\label{Chlorine}
\end{figure}

As mentioned in the main text, Fe(III)-porphine-chloride molecules are evaporated on a Pb(111) sample held below room temperature, followed by annealing the preparation to 370$\uK$. We show in Fig.~\ref{Chlorine}a a topography image obtained on a sample before annealing. There, we observe the formation of ordered islands in which a few molecules ($\leq 1 \%$) exhibit a pronounced protrusion above their center. As shown in Fig.~\ref{Chlorine}b, the apparent height of such a molecule is approximately $0.7\mathrm{~\AA}$ larger than the one of molecules with a clover shape. A similar elevation has been observed for Cl ligands on other Fe porphin derivatives, e.g., Fe-octaethylporphyrin-chloride (Fe-OEP-Cl) on Au(111)~\cite{SHeinrich2013} and Pb(111)~\cite{SHeinrich2013b}. The \didv spectra on these bright protrusions reveal the presence of sharp peaks outside the superconducting energy gap at bias voltages of $\pm 4\umV$ and $\pm 5.4\umV$ (see Fig.~\ref{Chlorine}c). These peaks reflect inelastic excitations of the molecule on a superconductor~\cite{SHeinrich2013b}, which can be assigned to spin excitations of the Fe center. In the presence of the Cl ligand the Fe center lies in $+3$ oxidation state with $S=5/2$. The anisotropic environment lifts the degeneracy of the $d$ levels. Together with spin-orbit coupling, this introduces magnetocrystalline anisotropy to the Fe states. The inelastic peaks in the \didv spectrum arise from transitions between these anisotropy-split levels. The absence/presence of these excitations thus indicates the absence/presence of the Cl ligand. All molecules discussed in the main text have lost their Cl ligand after annealing the deposited molecules to 370 $\uK$

\subsection{Quantum phase transition in molecules in different surroundings}

The results presented in the main text are acquired above a molecule at the border of an island. Here, we show that the same observations can be made for molecules in different surroundings. 

\begin{figure}[]
	\includegraphics [width=0.95\textwidth,clip=]{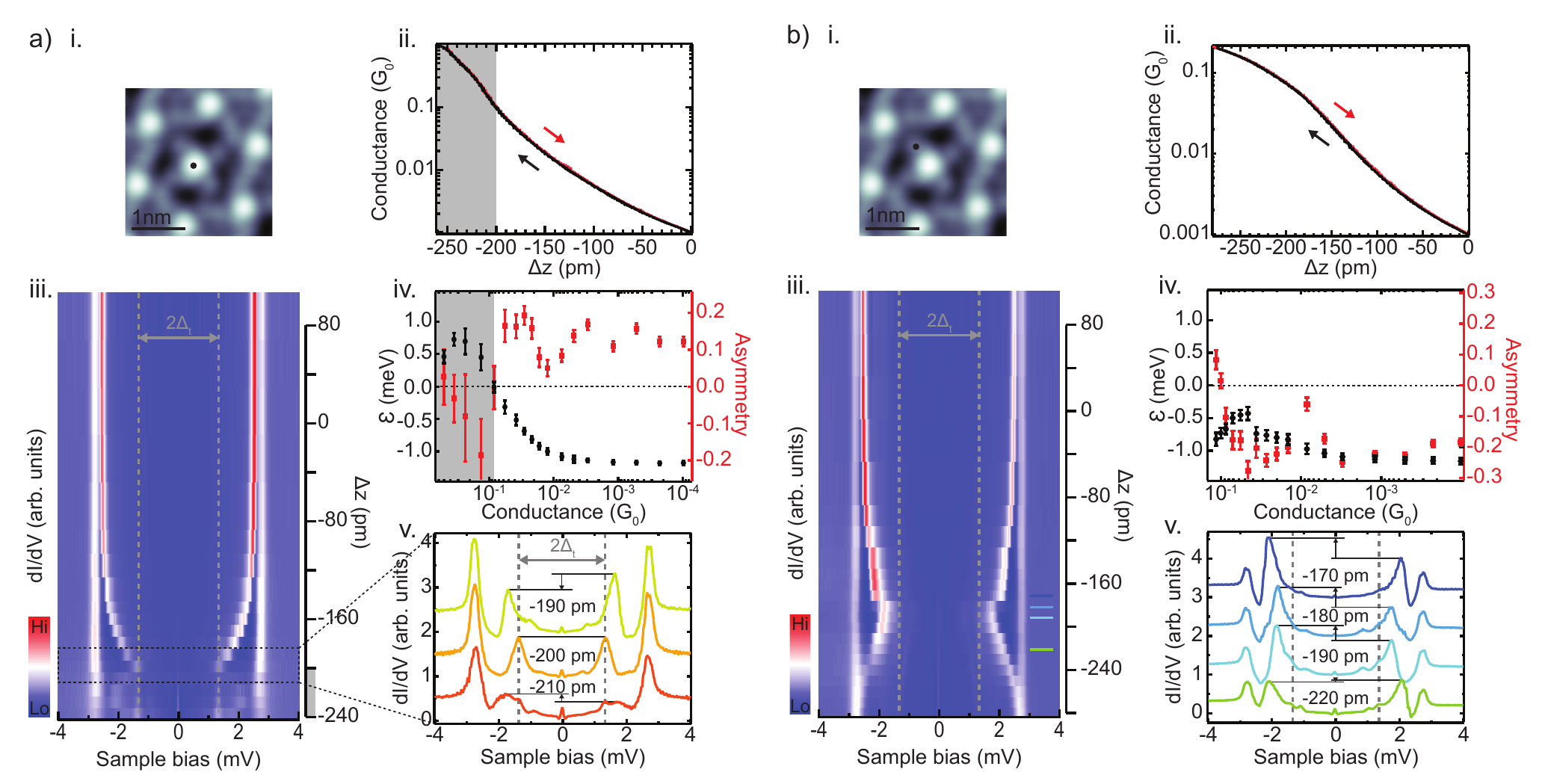}
	\caption{Evolution of the spectroscopic features above the center a) and ligand b) of a molecule within an island. i. STM topography image showing the tip location (black dot). ii. Dependence of the junction conductance upon tip approach. iii. Set of \didv spectra normalized to their conductance at 5$\umV$ recorded at different tip height. iv. Extracted YSR energy and asymmetry as a function of the junction conductance. v. Selected spectra (offset for clarity) showing that the system goes through the quantum phase transition on the center of the molecule (a), and does not go through the quantum phase transition by approaching on the ligand (b).}
	\label{QPT}
\end{figure}

The evolution of the junction conductance above the center and ligand of such a molecule within an island shows a reversible contact formation similar to the molecule in the main text [Fig.~\ref{QPT}(ii)]. A set of spectra recorded at different tip--sample distances is shown as a false 2D color plot in Fig.~\ref{QPT}(iii). At far tip  distance, a pair of YSR states is observed at $\pm 2.5\umV$ at both the center and the ligand of the molecule. At both locations the YSR states first shift toward zero energy upon tip approach, indicating a screened spin ground state and negative YSR energy. Above the center, the YSR state crosses zero energy and reverses its asymmetry around $\Delta z=-200\upm$ [see Fig.~\ref{QPT}a (iv and v)] indicating that the system undergoes the quantum phase transition. In contrast, above the ligand, the YSR energy remains negative and a reversal of the YSR asymmetry only occurs at high junction conductance ($G \geq 0.1 \times G_0$) when Andreev reflections dominate the transport through the junction [see Fig.\ref{QPT}b (iv and v)]. 

\begin{figure}[]
	\includegraphics [width=0.95\textwidth,clip=]{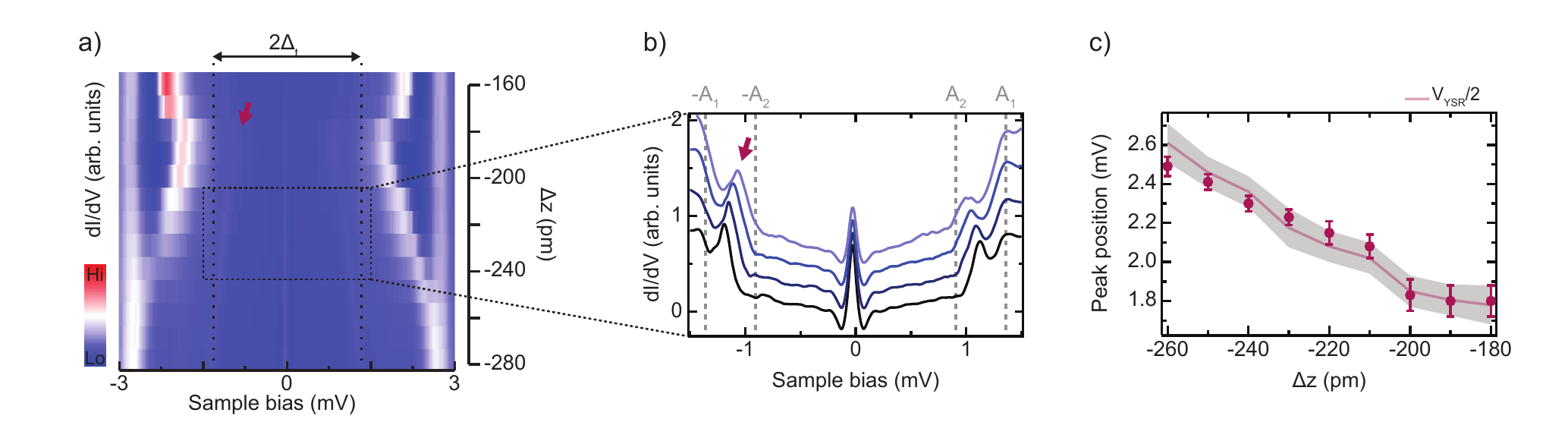}
	\caption{a) Close-up view of the approach set of Fig.~\ref{QPT}b highlighting the presence of resonances due to multiple Andreev reflections. b) Four spectra of this approach set (offset for clarity) showing a peak shifting upon tip approach. c) Extracted peak position as well as the expected position of the Andreev process according to the position of the YSR state.}
	\label{Andreev}
\end{figure}

As shown in Fig.~\ref{Andreev}, resonances within $V=\pm \Delta_t$ build up upon further tip approach. In addition to a zero-bias peak due to Cooper pair tunneling, resonances are observed at $V=\pm \frac{\Delta+\Delta_t}{2}$ and $V=\pm \frac{\Delta+\epsilon}{2}$ (see Fig.\ref{Andreev}c) arising from multiple Andreev reflections, as discussed in the main text.

\end{document}